\documentclass[12pt, draftclsnofoot, onecolumn]{IEEEtran}

\ifCLASSINFOpdf

\else

\fi

\usepackage{graphicx}
\usepackage{color}

\begin{document}

\title{Area Spectral Efficiency Analysis and Energy Consumption Minimization in Multi-Antenna Poisson Distributed  Networks}

\author{Zheng Chen,~
        Ling~Qiu,~\IEEEmembership{Member,~IEEE}, and Xiaowen Liang
        \thanks{The research has been supported by the 863 Program 2014AA01A702 and Specialized Research Fund for the Doctoral Program of Higher Education 20133402110061. The authors are with Key Laboratory of Wireless-Optical Communications, Chinese Academy of Sciences, School of Information Science and Technology, University of Science and Technology of China (email: lygchenz@mail.ustc.edu.cn; lqiu@ustc.edu.cn; lxw@ustc.edu.cn).  L. Qiu is the corresponding author.}}

\markboth{}%
{Shell \MakeLowercase{\textit{et al.}}: Bare Demo of IEEEtran.cls for Journals}

\maketitle

\begin{abstract}
This paper aims at answering two fundamental questions: how area spectral efficiency (ASE) behaves with different system parameters; how to design an energy-efficient network. Based on stochastic geometry, we obtain the expression and a tight lower-bound for ASE of Poisson distributed networks considering multi-user MIMO (MU-MIMO) transmission. With the help of the lower-bound, some interesting results  are observed. These results are validated via numerical results for the original expression. We find that ASE can be viewed as a concave function with respect to the number of antennas and active users. For the purpose of maximizing ASE, we demonstrate that the optimal number of active users is a fixed portion of the number of antennas. With optimal number of active users, we observe that ASE increases linearly with the number of antennas. Another work of this paper is joint optimization of the base station (BS) density, the number of antennas and active users to minimize the network energy consumption. It is discovered that the optimal combination of the number of antennas and active users is the solution that maximizes the energy-efficiency. Besides the optimal algorithm, we propose a suboptimal algorithm to reduce the computational complexity, which can achieve near optimal performance.
\end{abstract}


\IEEEpeerreviewmaketitle

\section{Introduction}

\IEEEPARstart{N}{etwork} densification is  a key  approach to cope with the 1000x traffic demand of 5G cellular networks \cite{dense:bushan}\cite{metis:d11}. Especially, spatial densification,  which includes increasing the number of antennas  per BS (i.e., multi-input multi-output, MIMO) and the number of BSs (i.e., ultra dense network, UDN), has been recognized as an essential part of future 5G cellular networks.

In this context, a fundamental question arises, which is how the network throughput behaves with the BS density, the number of antennas and other system parameters. This is of crucial importance since we should know the appropriate amount of infrastructure (e.g., BSs and antennas per site) that can  meet the network throughput demands.  On the other hand, energy-efficiency is also an important requirement of 5G cellular network, where the energy-efficiency should be improved at least 100x \cite{5g:andrews}. Specifically, as increasing the BS density and the number of antennas both boost the network capacity, it is interesting to find  an optimal combination of BS density and number of antennas for reducing network energy consumption.

To address the above two fundamental questions, we need an efficient approach to evaluate the performance of the whole network. Previous studies on the network throughput or energy-efficiency mainly focused on the hexagonal model, in which the macro cells are modeled as hexagonal grids and the small cells are distributed in each macro cell \cite{3gpp:model}. Due to the  irregularity of future small cells, the regular grid models may be too idealized to characterize the network performance accurately. Furthermore, such works highly rely on the time-consuming system-level simulations, which makes it difficult to shed more lights on system design.   For mathematical tractability, stochastic geometry has been applied to the analysis and optimization of cellular networks in recent years \cite{elsawy:survey}.

\subsection{Related Works}
Stochastic geometry has been used for the analysis of cellular networks since late 90's. \cite{baccelli:sg} introduced stochastic geometry as a communication network planning tool. By modeling the BSs as a Poisson point process (PPP), \cite{brown:shotgun} studied the signal-to-interference-plus-noise-ratio (SINR) of the cellular networks. Andrews \emph{et al}. derived more tractable expressions for SINR and average data rate in \cite{homogeneous:andrews}. Extensive work on heterogeneous networks (HetNets) can be found in \cite{hetnet:jo}. Some interesting results have been obtained in \cite{homogeneous:andrews}\cite{hetnet:jo}. For example, without considering cell range expansion and the thermal noise, the SINR distribution of users does not depend the transmit power of BSs nor the BS densities. In order to better suit reality networks, more complex point processes have also been adopted in  the subsequent studies. Considering the fact that there exists repulsion between the well-planned macro BSs, \cite{deng:gpp} modeled the cellular networks   as  Ginibre point process (GPP) and analyzed the mean interference and coverage probability. On the other hand, for the hot areas where lots of small cells are deployed, Poisson cluster process (PCP) can be used to model the BSs with attraction \cite{vanay:pcp}.

Based on the analysis from stochastic geometry, some other works focused on the optimal system parameters to improve the network performance. Especially, from the perspective of energy-efficient networks, the optimal deployment strategies have been widely studied. As the BSs are reported to consume 60-80\% energy of the whole network \cite{marsan:6080}, most researchers mainly evaluated the energy consumption of BSs. Based on stochastic geometry, \cite{peng:trans} jointly optimized the transmit power of BSs and BS densities  under coverage performance constraint.  Besides the coverage performance, improving each user' data rate is also a primary goal of future networks. In this context, \cite{energy:cao} took the load of each BS into consideration  and studied the energy optimal BS densities while guaranteeing users' data rate constraint. Considering the huge energy waste of UDN in the off-peak period, BS sleeping is an efficient approach to reduce the network energy consumption. To characterize the potential gains of BS sleeping, the performance of different sleeping strategies has been analyzed in \cite{energy:soh}. Moreover, to support flexible management, the introduction of UDN also brings about changes in network architecture. Under the recently proposed separation architecture, \cite{wang:separation} derived the optimal small cell density considering both the coverage and data rate constraints.

However, most of the afore-mentioned works assumed only single antenna is equipped for each BS, i.e., the potential gains of multi-antenna cannot be revealed. To take a step further, the coverage probability and average data rate in Poisson distributed network with multi-antenna BSs have been obtained in \cite{dhillon:mimohetnet}. Based on the approach of equivalent-in-distribution approach, \cite{marco:mimo} studied the error probability of Poisson distributed networks with different MIMO arrangements. Due to the intractability  of the expression for the Laplace function of the inter-cell interference, it is difficult to study the relationship between network performance and system parameters (e.g., the number of antennas) in \cite{dhillon:mimohetnet}\cite{marco:mimo}. Hence, with the help of Toeplitz matrix, \cite{li:eemimo} derived a more tractable expression for the coverage probability considering maximal ratio transmission (MRT) beamforming.  It is found that the benefit of adding more antennas will become smaller with the increase of the number of antennas. Extension of  \cite{li:eemimo} can be found in \cite{li:sdma}, in which space-division multiple access (SDMA) was studied. Although \cite{li:eemimo}\cite{li:sdma} have provided some interesting results, the limitation is that they only considered fixed-rate transmission. This may be not very suitable for today's cellular network, where adaptive modulation and coding (AMC) is applied (i.e., average data rate may be more appropriate).  In future networks, a large number of antennas might be deployed, which is referred as massive MIMO. Assuming the number of antennas is sufficiently large, \cite{bai:massive}\cite{bjornson:massive} studied the performance of massive MIMO networks. Furthermore, the system parameters were jointly optimized to maximize the energy-efficiency in \cite{bjornson:massive}.

In this paper, we aim at fully characterizing the properties of the network throughput and energy consumption in Poisson distributed networks with multi-antenna BSs. We consider MU-MIMO in this paper, i.e., more than one users will be scheduled in each slot each cell. Specifically, zero-forcing (ZF) precoding is adopted  to eliminate the intra-cell interference. Based on the theory of stochastic geometry, we derive the expression and a lower-bound for the average data rate of a typical user. Then,  ASE of the network can be expressed with these two expressions. Based on the analysis results,  how ASE behaves with other system parameters (i.e., the BS density, the number of antennas, and the number of active users) is discussed in detail.  Thus, the potential gains of more BSs and more antennas are characterized.   Compared to previous work, more interesting and insightful results have been obtained. Especially, the functional relationship between ASE and system parameters (i.e., the number of antennas, the number of active users) is fully quantified.   Furthermore, to minimize the network energy consumption,  we formulate an optimization problem, where the BS density, the number of antennas, and the number of active users are jointly optimized. This work provides system design insight for the deployment of future energy-efficient network.

\subsection{Contributions}
In this paper, the functional relationship between ASE and other system parameters will be studied theoretically. From the perspective  of ASE and network energy consumption, the optimal configurations of system parameters will be discussed. The main contributions of this paper are summarized as follows:
\begin{itemize}
  \item Using tools from stochastic geometry, we derive the expression and a lower-bound for the average data rate of a typical user. Numerical results show that the low-bound is quite tight in most cases. Additionally, the lower-bound is quite useful in the analysis and optimization of this paper. We discover that the average data rate mainly depends on the ratio between the number of active users and the number of antennas.
  \item The functional relationships between ASE and other system parameters are discussed. It is observed that ASE increases linearly with the BS density under the infinite user density assumption. Another result that has never been reported in previous works is that, ASE is a concave function with respect to the number of antennas and active users when treating them as continuous variables.
  \item For arbitrary number of antennas, we demonstrate that the optimal number of active users to maximized ASE is approximately a fixed portion of the number of antennas. With the number of active users set as the optimal number, we find that ASE increases linearly with the number of antennas. We define GAPA as the gain on ASE per antenna to quantify the performance gain by deploying one more antenna.
  \item Although the above results are mostly obtained from the more tractable lower-bound. These properties of the original expression are validated through numerical results. Because of the tightness of the lower-bound, we find the properties of the original expression are consistent with those of the lower-bound.
  \item To pursue the energy optimal deployment strategy, we jointly optimized the system parameters to minimize the network energy consumption. We find that the optimal combination of the number of antennas and number of active users is the solution that maximizes the energy-efficiency. Besides the optimal algorithm, we also proposed a suboptimal algorithm based on  the lower-bound to solve the non-convex problem. Numerical results show that the performance loss of the suboptimal algorithm is negligible. And we find that we should equip more antennas and schedule more active users for the BSs which have higher transmit power.
\end{itemize}

\subsection{Paper Organization}
The rest of this paper is organized as follows. Section \ref{secsystemmode} describes the system model and performance metrics in this paper. Section \ref{secanalysis} analyzes ASE of the whole network. In Section \ref{secimpact}, we discuss the relationship between ASE and other system parameters. The system parameters are jointly optimized in Section \ref{secoptimal}.  Finally, Section \ref{secconclusion} summarizes and concludes this paper.

\emph{Notation}: We use underline $\underline{(\cdot)}$ to indicate the results obtained based on the lower-bound. We use the subscripts $(\cdot)_{ASE}$ and $(\cdot)_{EC}$ to indicate the results related to ASE maximization and energy consumption minimization, respectively.

\section{System Model and Performance Metrics}\label{secsystemmode}

\subsection{Network Model}

\begin{figure}[!t]
\centering
\includegraphics[width=3.2in]{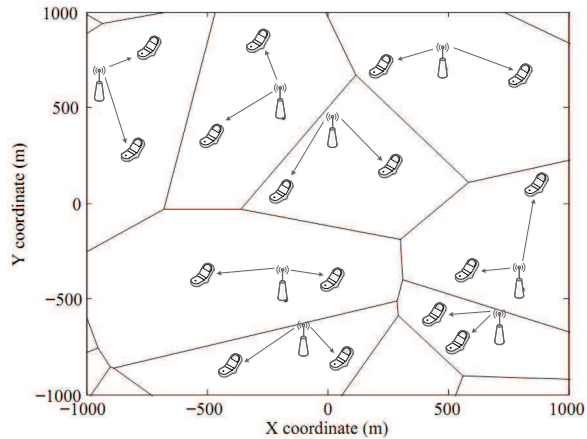}
\caption{Illustration of the network model.}
\label{topo}
\end{figure}

We consider a downlink cellular network, which is depicted in Fig. \ref{topo}. The BSs are located according to a homogeneous PPP $\Phi_b=\left\{x_0,x_1,...\right\}$ in the Euclidean plane. The intensity of $\Phi_b$ is $\lambda_b$. We assume that each BS is equipped with $M$ antennas. The single-antenna users are arranged according to some stationary point process. The users' point process is independent of the base stations' point process. Each user is associated with the nearest BS, i.e., each BS serves the users which are located within its Voronoi cell. We apply standard path loss propagation model with path loss exponent  $\alpha>2$. Since cellular networks are typically interference limited \cite{interference:limited}, we neglect the effect of the thermal noise in this paper. The small scale fading on each link is i.i.d. Rayleigh fading. The correlations between different antennas are ignored. Due to the limitation of frequency resource, we assume universal frequency reuse is applied in this paper.

At  each time slot, each BS serves $K$ users through SDMA, i.e., the number of active user is $K$. We assume the user density is larger enough that there are at least $K$ users within each BS's Voronoi cell, which is a common assumption in previous works \cite{dhillon:mimohetnet}\cite{li:sdma}. Besides, this is a reasonable assumption in 5G cellular network, since to provide service for dense crowds of users is a typical scenario of METIS \cite{metis:d11}. In this paper, we mainly focus on ZF precoding and assume perfect channel state information at each BS, i.e., there is no intra-cell interference. Cooperation between BSs is not considered in this paper. As a result, the number of active users $K$ should not exceed the number of antennas, i.e., $K \le M$. The total transmit power of each BS is $P$. We consider equal power allocation for the $K$ users in each cell.

Let the $M\times1$ vector $\mathbf{h}_{ik} \sim \mathcal{CN}(0,\mathbf{I})$ denote the small scale fading between the $i$-th BS and its $k$-th active user. Let $\mathbf{w}_{ik}$ denote the precoding vector at BS $i$ for the $k$-th user. Considering ZF precoding, the unit-norm precoding vector $\mathbf{w}_{ik}$ equals the $k$-th normalized column of $H_i\left(H_i^{\dag}H_i\right)^{-1}$, where $H_i=\left[\mathbf{h}_{i0},\mathbf{h}_{i1},...,\mathbf{h}_{i(K-1)}\right]$. The transmitted signal of the $i$-th BS is $\mathbf{z}_i=\sum_{k=0}^{K-1} \mathbf{w}_{ik}s_{ik}$, where $s_{ik}$ is the data symbol destined for the $k$-th active user. Without loss of generality, we assume the $0$-th active user of the $0$-th BS is located at the origin, which is referred as the typical user \cite{homogeneous:andrews}. Based on Slivnyak¡¯s Theorem \cite{book:baccelli}, we mainly focus on the analysis of the typical user. The small scale fading between the interfering BSs and the typical user is denoted as $\mathbf{v}_{i0}\sim \mathcal{CN}(0,\mathbf{I})$. The received signal of the typical user is
\begin{equation}\label{eqsignal}
y_0 = \sqrt{\frac{P}{K}}\left\|x_0 \right\|^{-\frac{\alpha}{2}}\mathbf{h}_{00}^{{\dag}}\mathbf{z}_0+ \sum_{x_i \in \Phi_b \backslash \{x_0\} } \sqrt{\frac{P}{K}} \left\|x_i\right\|^{-\frac{\alpha}{2}}\mathbf{v}_{i0}^{{\dag}}\mathbf{z}_i
\end{equation}
From $(\ref{eqsignal})$, the signal-to-interference-ratio (SIR) of the typical user can be expressed as
\begin{equation}\label{eqsir}
SIR_0=\frac{g_{00}\left\|x_0 \right\|^{-\alpha}}{\sum_{x_i \in \Phi_b \backslash \{x_0\} }g_{i0}\left\|x_i\right\|^{-\alpha}},
\end{equation}
where $g_{00}=\left\|\mathbf{h}_{00}^{\dag}\mathbf{w}_{00}\right\|^2$ and $g_{i0}=\sum_{k=0}^{K-1}\left\|\mathbf{v}_{i0}^{\dag}\mathbf{w}_{ik}\right\|^2$ for $i>0$. The desired channel power $g_{00}$ is the squared-norm of the projection of vector $\mathbf{h}_{00}$ on  $\textrm{Null}\left(\mathbf{h}_{01},...,\mathbf{h}_{0(K-1)}\right)$, which follows $Gamma(M+1-K,1)$ \cite{jindal:simo}. For the interfering links, as $\mathbf{w}_{ik}$ is a unit-norm vector and independent of $\mathbf{v}_{i0}^{\dag}$, $\left\|\mathbf{v}_{i0}^{\dag}\mathbf{w}_{ik}\right\|^2$ is the squared-norm of complex Gaussian, which is exponential distributed. We follow the approximation in \cite{dhillon:mimohetnet}, i.e., neglect the correlation between $\left\|\mathbf{v}_{i0}^{\dag}\mathbf{w}_{ik}\right\|^2$ for different $k$. Therefore,  the equivalent channel gain $g_{i0}$ is the sum of $K$ independent exponential distributed random variables, which follows $Gamma\left(K,1\right), i>0$. The accuracy of this approximation will be demonstrated in our numerical results.

\subsection{Performance Metrics}
Throughput and energy-efficiency are two key performance indicators of future cellular networks \cite{metis:d11}.  Therefore, we will focus on the performance metrics of ASE and network energy consumption in this paper.

\subsubsection{Area Spectral Efficiency}
One of the major purpose of this paper is to study the relationship between ASE and other system parameters (e.g., $P,\lambda_b, M, K$). ASE is  the performance metric reflecting the network capacity, which is defined as the average throughput per HZ per unit area. In accordance with \cite{li:eemimo}\cite{li:sdma}\cite{bjornson:massive}, ASE is the product of BS density, the number of active users served by each BS, and the average data rate of the typical user. To reveal the potential gains via spatial densification, the functional relationship between ASE and other system parameters is of crucial importance. Through the expressions of ASE, the theoretical gains of increasing BS density and number of antennas will be characterized in this paper. Some interesting and insightful observations will be obtained.

\subsubsection{Network Energy Consumption}
Future cellular networks should meet the explosive demand of data rate at the cost of similar energy consumption with concurrent networks. Thus, network energy consumption is another focus of this paper. Since the BSs consume the largest portion of energy in cellular networks, we will mainly evaluate the overall BS energy consumption. For each BS, we adopt the following energy consumption model \cite{zhang:model}\footnote{We do not consider the energy consumption of backhaul in this paper. Thus, the energy consumption model is slightly different from \cite{zhang:model}. However, the results in this paper can be applied to the scenario considering backhaul energy consumption directly.}
\begin{equation}
EC=\frac{P}{\eta}+MP_c+K^3P_{pre}+P_0,
\end{equation}
where $\eta$ denotes the power amplifier efficiency, $P_c$ is the circuit power per antenna, which indicates the energy consumption of the corresponding RF chains. $K^3P_{pre}$ accounts for the energy consumption for precoding which is related to the number of active users.  $P_{0}$ is the non-transmission power, which  accounts for the energy consumption of baseband processing, cooling, etc. Considering the BS energy consumption model, network energy consumption is defined as the average energy consumption per unit area, i.e.,
\begin{equation}
NEC=\lambda_b\left(\frac{P}{\eta}+MP_c+K^3P_{pre}+P_0\right).
\end{equation}

In the following part of this paper, using stochastic geometry, we will discuss the relationship between ASE and other system parameters. Furthermore, given an ASE target, how to design an energy optimal network will also be studied.

\section{Analysis of  Area Spectral Efficiency}\label{secanalysis}
In this section, we will first derive the expression for the average data rate of the typical user. Then we will give the expression of ASE. Besides, to further study how ASE behaves with  different system parameters, a more tractable lower-bound  will also be provided, which is demonstrated to be quite tight through numerical results.

\subsection{Area Spectral Efficiency}
ASE is the product of BS density $\lambda_b$, the number of active users $K$ and average data rate of users $E\left[R\right]$. Hence, we first need to derive $E\left[R\right]$ to get the expression of ASE.  Similar to \cite{marco:datarate}, we derive the average data rate of the typical user directly without calculation of SIR distribution. And more tractable expressions are provided. The tractability facilitates the analysis of properties of ASE in the following section.

\newtheorem{theorem}{Theorem}
\begin{theorem}\label{thase}The average data rate of the typical user is given by
\begin{equation}\label{eqase}
E[R]=\int_0^\infty  \frac{ {\frac{{1}}{z}} \left({1 - \left( {\frac{1}{{1 + z}}} \right)^{M + 1 - K} }\right)}{{\frac{1}{\left( {1 + z} \right)^{K}}  + z^{\frac{2}{\alpha }} KB\left( {\frac{z}{{1 + z}}, \frac{\alpha-2}{\alpha},K + \frac{2}{\alpha} } \right)}}dz,
\end{equation}
where $B\left( \cdot,\cdot,\cdot \right)$ is the incomplete Beta function.
\end{theorem}

\begin{IEEEproof}
See Appendix \ref{proofase}.
\end{IEEEproof}

Theorem \ref{thase} provides an expression for the average data rate of users. An observation from Theorem \ref{thase} is that, the average data rate of the typical user  does not depend on the BS density $\lambda_b$ nor the transmit power $P$. This phenomenon is similar to the single antenna case in \cite{homogeneous:andrews}. Nevertheless, we consider a more general case, where  multi-antenna and MU-MIMO are both taken into account. It is important to point out that, the density and transmit power invariance property does not hold when considering multi-tier network. It has been shown that the SIR distribution depends on the BS density and transmit power of different tiers in \cite{li:sdma}. However, the emphasis of this paper is the analysis and optimization of single-tier network. Further extension on multi-tier network is left for future works. Based on Theorem \ref{thase}, ASE of the network can be expressed as
\begin{equation}\label{eqt}
T=\lambda_bKE\left[R\right].
\end{equation}

Although a tractable expression for $E\left[R\right]$ is derived,  we also find that it is cumbersome to study the relationship between $E\left[R\right]$ and the number of active users $K$. Therefore, we also derive a tight lower-bound for $E\left[R\right]$, which will be presented in Theorem \ref{thlowbound}.

\subsection{A Tight Lower-Bound}
To shed more lights on the system design of future network, we obtain a tight lower-bound $\underline{E\left[R\right]}$ for the average data rate of the typical user.

\begin{theorem}\label{thlowbound}A lower-bound for the average data rate of the typical  user  is given by
\begin{equation}\label{eqlowebound}
\underline{E\left[R\right]}=\int_0^\infty  {\frac{{1}}{z}} \frac{{1 - e^{ - z\frac{{M - K}}{K}} }}{{ e^{ - z}  + z^{\frac{2}{\alpha }} \gamma \left( {1 - \frac{2}{\alpha },z} \right)}}dz,
\end{equation}
where $ \gamma \left( \cdot,\cdot \right)$ is the lower incomplete Gamma function.
\end{theorem}

\begin{IEEEproof}
See Appendix \ref{prooflowebound}
\end{IEEEproof}


\begin{figure}[!t]
\begin{minipage}[t]{0.5\linewidth}
\centering
\includegraphics[width=3.2in]{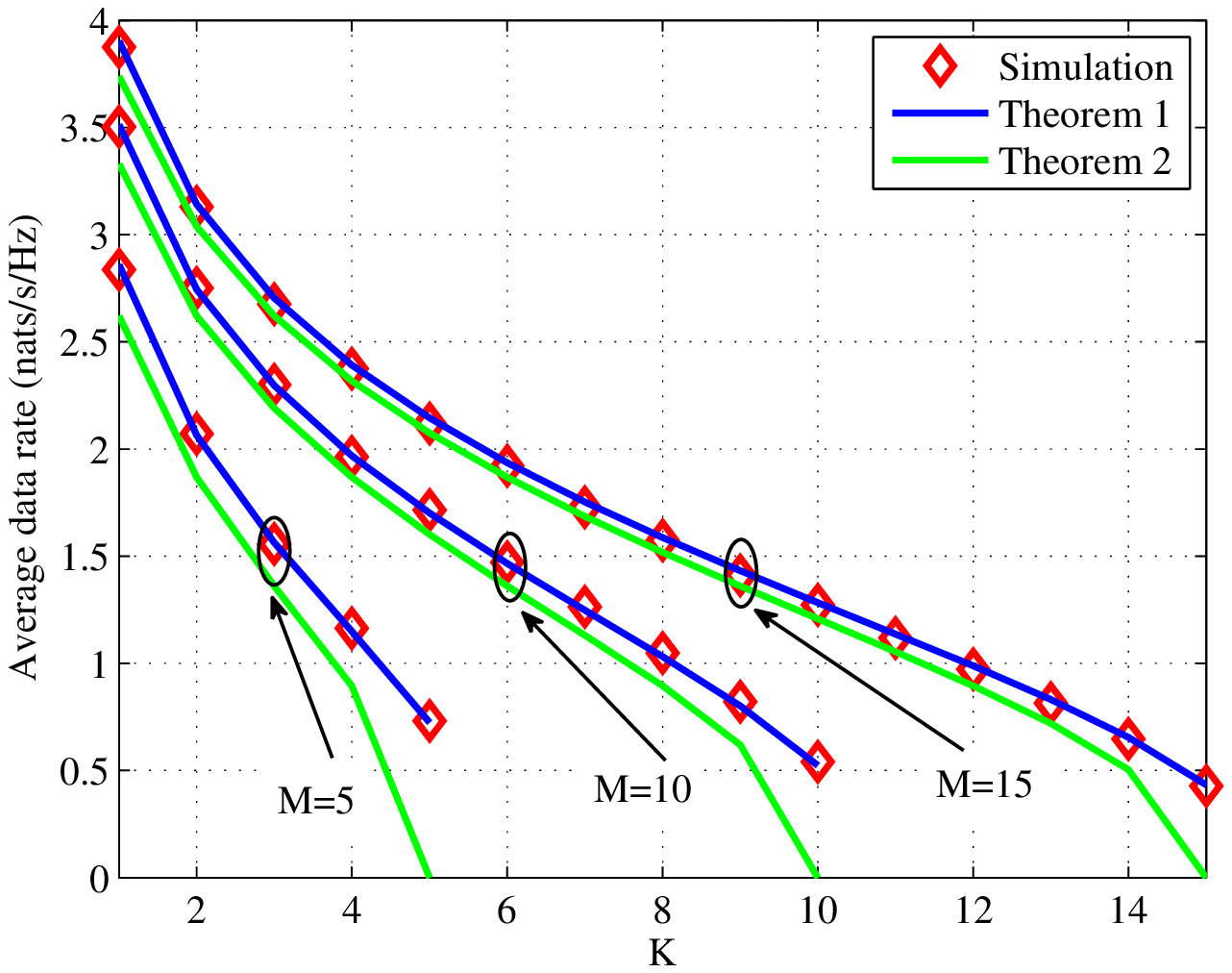}
\caption{Average data rate  with different $K$. $\alpha=4$.}
\label{datarate}
\end{minipage}
\hspace{1ex}
\begin{minipage}[t]{0.5\linewidth}
\centering
\includegraphics[width=3.2in]{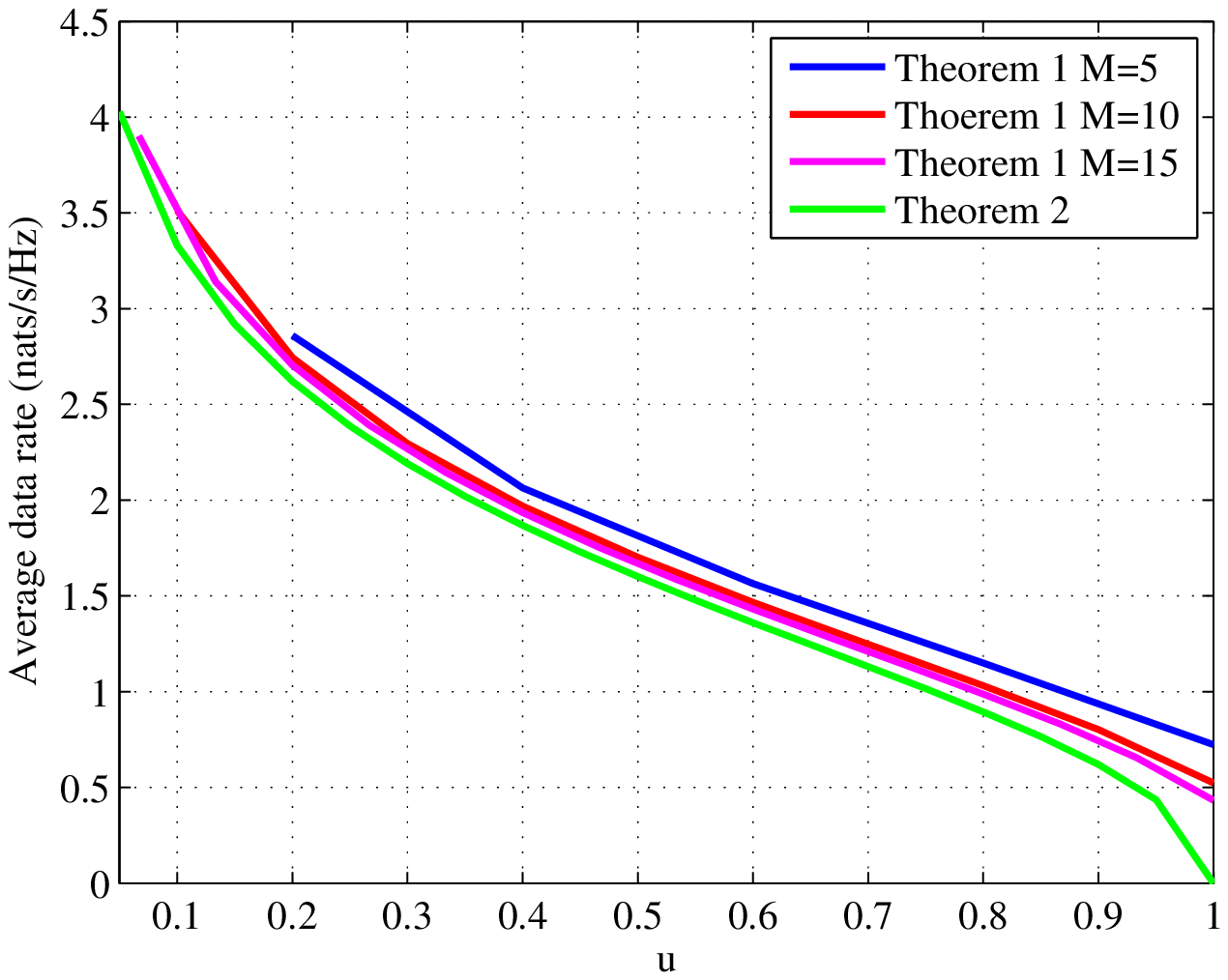}
\caption{Average data rate  with different $u$. $\alpha=4$.}
\label{datarateu}
\end{minipage}
\end{figure}

Similar to Theorem \ref{thase}, the lower-bound in Theorem \ref{thlowbound} does not depend on $P$ nor $\lambda_b$. Fig. \ref{datarate} compares the results of Monte Carlo simulations, Theorem \ref{thase} and Theorem \ref{thlowbound}. Firstly, we can find that the analytic results in  Theorem \ref{thase} match the simulation results quite well.\footnote{In order to validate the approximation for the channel power of the interfering links,  the values of $g_{i0}$ are calculated based on the precoding matrices in simulations. From the numerical results, we find that Gamma approximation is quite accurate for the analysis of average data rate.} Secondly, the lower-bound proposed in Theorem \ref{thlowbound} is quite tight for $K<M$, especially when the number of antennas is large. Besides, the average data rate decreases with the number of active users $K$. The reason is that, if more users are scheduled in each cell, the degrees of freedom to boost the received useful power will be reduced while the inter-cell interference will be stronger. In fact, it is not difficult to show that $\underline{E\left[R\right]}$ decreases with $K$ from Theorem \ref{thlowbound}. However, it is cumbersome to derive  this property based on Theorem \ref{thase} theoretically. In other words, the tight  lower-bound $\underline{E\left[R\right]}$ can provide an efficient way to study the relationship between the network performance and other system parameters theoretically.


From Theorem \ref{thlowbound}, we find that the lower-bound $\underline{E\left[R\right]}$ only depends on the ratio between the number active users and the number of antennas $u=\frac{K}{M}$. Fig. \ref{datarateu} illustrates the average data rate of the typical user with different $u$. It is worth noting that, even for the expression in Theorem \ref{thase}, $E\left[R\right]$ is mainly related to the ratio $u$. The mismatch of considering different $M$ is relatively small. This phenomenon can be explained by averaging the equivalent channel gains. Similar to the procedures in Appendix B, taking expectations of  $\frac{1}{g_{00}},g_{i0}$, we have $SIR_0 \approx \frac{\left(M-K\right)\left\|x_0 \right\|^{-\alpha}}{K\sum_{x_i \in \Phi_b \backslash \{x_0\} }\left\|x_i\right\|^{-\alpha}}= \left(\frac{M}{K}-1\right)\frac{\left\|x_0 \right\|^{-\alpha}}{\sum_{x_i \in \Phi_b \backslash \{x_0\} }\left\|x_i\right\|^{-\alpha}}$. We can see that the approximate users' SIR only depends on the ratio $u$. Therefore, we observe the results on average data rate in Fig. \ref{datarateu}.

 From Theorem \ref{thlowbound}, the lower-bound  of ASE is given by
\begin{equation}\label{lowboundt}
\underline{T}=\lambda_bK\underline{E\left[R\right]}.
\end{equation}
Based on $(\ref{eqt})$ and $(\ref{lowboundt})$, more detailed discussions on the variation trends of ASE with different system parameters will be provided in the next section, where Theorem \ref{thlowbound} also plays a crucial rule in theoretical analysis.

\section{Impact of System Parameters on ASE}\label{secimpact}
In the previous section, we have  provided the expression and a lower-bound for ASE.  In this section, we will further discuss how ASE behaves with different system parameters. The following analysis provides system design insight on how to design an efficient network. We mainly focus on three system parameters, i.e., the BS density $\lambda_b$, the number of antennas $M$ and  the number of active users $K$.

\subsection{Relationship Between ASE and Other System Parameters}
\subsubsection{BS Density}
As $E\left[R\right]$ is only related to $M$ and $K$,  ASE $T$ increase linearly with the BS density $\lambda_b$ when $M,K$ are fixed. That is to say, deploying more BSs is an efficient way to boost the network capacity. It is worth noting that, this conclusion only holds  under the assumption that the user density  is always sufficiently larger than the BS density, i.e., the user density is infinite.If the user density is not so large (e.g., rural areas), the relationship between ASE  and BS density can be investigated similarly to \cite{li:eemimo}. Nevertheless, this is not the emphasis of this paper, more detailed discussion on BS density when the user density is not sufficiently large will be left for future works.

\subsubsection{Number of BS Antennas}
If the BS density $\lambda_b$ and the number of active users are fixed, we have the following results on the number of antennas.

\newtheorem{proposition}{Proposition}
\begin{proposition}\label{m}If the BS density $\lambda_b$ and the number of active users $K$ are fixed, ASE $T$ is an increasing concave function of the number of antennas $M$.\footnote{In this section, we treat $M,K$ as continuous variables to study the properties of ASE.}
\end{proposition}

\begin{IEEEproof}We need to prove that $E\left[R\right]$ is  an increasing concave function of $M$. The only item related to $M$ in $E\left[R\right]$ is $-\left( {\frac{1}{{1 + z}}} \right)^{M + 1 - K} $ . It is not difficult to show that  $-\left( {\frac{1}{{1 + z}}} \right)^{M + 1 - K} $ is an increasing and concave function of $M$. Combining the fact that integration is a linear operation, we know that $E\left[R\right]$ is also an increasing concave function of $M$.
\end{IEEEproof}

From Proposition \ref{m}, we know that deploying more antennas always improves the network capacity. However, the benefit of more antennas will become smaller with the increase of the number of antennas. Hence, it seems like that increasing the number of antennas is not such attractive. Nevertheless, our following analysis shows that if the number of active users can be adjusted adaptively to $M$, more gains on ASE can be achieved.

\subsubsection{Number of active users}
The number of active users $K$ affects both the power distribution of desired signal and interfering signal. It is of vital  important to study how the number of active user affects the network capacity. However, the expression of $E\left[R\right]$ is complex with respect to $K$. Therefore, the analysis is based on the lower-bound $(\ref{lowboundt})$.

\begin{proposition}\label{prok}With the BS density $\lambda_b$ and the number of antenna $M$ fixed, the low-bound of ASE $\underline{T}$ is a concave function of the number of active users $K$.
\end{proposition}

\begin{IEEEproof}
The low-bound is expressed as
\begin{equation}\label{eqtmk}
\underline{T}=\lambda_b\int_0^\infty  {\frac{{1}}{z}} \frac{K\left({1 - e^{ - z\frac{{M - K}}{K}} }\right)}{{e^{ - z}  + z^{\frac{2}{\alpha }} \gamma \left( {1 - \frac{2}{\alpha },z} \right)}}dz.
\end{equation}
For the item $K\left({1 - e^{ - z\frac{{M - K}}{K}} }\right)$, we first get the second order derivative with respect to $K$. We have $\frac{\partial ^2}{{\partial K^2}}\left(K\left({1 - e^{ - z\frac{{M - K}}{K}} }\right)\right)=-\frac{z^2M^2}{K^3}e^{-z\frac{M}{K}+z}<0$.  Combining the fact that integration is a linear operation,  we know $\underline{T}$ is concave with respect to $K$.
\end{IEEEproof}

Since the low-bound of ASE $\underline{T}$ is a concave function of $K$, there exists a unique optimal $\underline{K}_{ASE}^*$ to maximize ASE $\underline{T}$.  The optimal number of active users will be discussed in detail in subsection B, in which some interesting results will be observed. Although the analysis is based on the low-bound $\underline{T}$,  numerical results in subsection B demonstrate that the analysis is consistent with the expression $T$.

\subsubsection{ASE with respect to $(M,K)$}
In the previous parts, we have discussed the functional relationship between ASE and $\lambda_b,M,K$ when the other two parameters are fixed. Herein, we will discuss the functional property of ASE with respect to the pair $(M,K)$. Although the analysis is based on $\underline{T}$, numerical results show that the conclusion also holds for the expression $T$.

\begin{theorem}\label{thmk}
The low-bound of ASE $\underline{T}$ is concave with respect to $\left(M,K\right)$.
\end{theorem}

\begin{IEEEproof}
See Appendix \ref{proofconcave}.
\end{IEEEproof}

\begin{figure}[!t]
\centering
\includegraphics[width=3.2in]{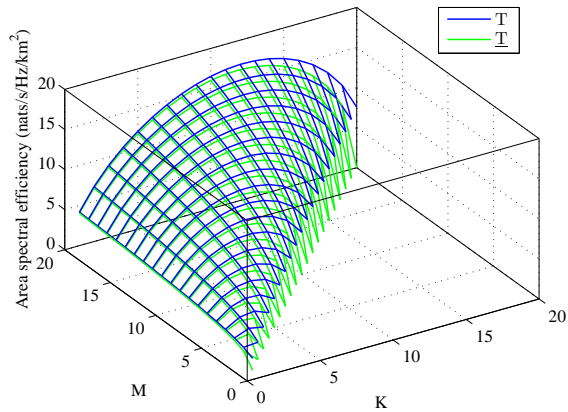}
\caption{ASE with $(M,K)$. $\lambda_b=1km^{-2}$. $\alpha=4$.}
\label{throughputall}
\end{figure}

Theorem \ref{thmk} reveals the functional relationship between ASE and $(M,K)$. To our best of knowledge, similar results have never been reported. Fig. \ref{throughputall} depicts $T$ and $\underline{T}$ with respect to $(M,K)$. It is intuitive that, $\underline{T}$ is concave with respect to $(M,K)$, which demonstrates our analysis. Furthermore, we find the original expression $T$ can also be viewed as  a concave function of $(M,K)$. That is to say the analysis based on the lower-bound $\underline{T}$ still holds for the expression $T$. Besides,  the results of Proposition \ref{m} and Proposition \ref{prok} can be viewed as  corollaries of Theorem \ref{thmk}.

\subsection{The Optimal Number of Active Users}
As we have analyzed, if given $M$, there exists an optimal active users $K_{ASE}^*$ to maximize ASE.  Therefore, we will study the optimal $K_{ASE}^*$ in this part. The following analysis is also based on the lower-bound $\underline{T}$. The optimal number of  active users derived based on the lower-bound is denoted as $\underline{K}_{ASE}^*$. Numerical results demonstrate that the results still hold when considering the expression $T$.
\subsubsection{ASE maximization problem}
 To derive $\underline{K}_{ASE}^*$, we formulate an optimization problem as
\begin{equation}
\begin{array}{lcl}
\mathbf{P1:} &     \max \limits_{K}   & \lambda _b K\int_0^\infty  {\frac{{1}}{z}} \frac{{1 - e^{ - z\frac{{M - K}}{K}} }}{{ e^{ - z}  + z^{\frac{2}{\alpha }} \gamma \left( {1 - \frac{2}{\alpha },z} \right)}}dz \\
&s.t. &   K \in \{1,2,3,...,M\}.
\end{array}
\end{equation}

Substituting $u=\frac{K}{M}$ into $\mathbf{P0}$, and relaxing $u$ to $[0,1]$, we have the following optimization problem.
\begin{equation}\label{problemt}
\begin{array}{lcl}
&     \max \limits_{u}   & \lambda _b M \underbrace{u\int_0^\infty  {\frac{{1}}{z}} \frac{{1 - e^{-z\frac{1}{u}+z} }}{{e^{ - z}  + z^{\frac{2}{\alpha }} \gamma \left( {1 - \frac{2}{\alpha },z} \right)}}dz}_{G(u)} \\
&s.t. &   u \in [0,1].
\end{array}
\end{equation}

Interestingly, from the optimization problem $(\ref{problemt})$, we know that the optimal $u_{ASE}^*$ is unrelated to $M$. That is to say, the ratio between the optimal number of active users and the number of antennas remains nearly the same for arbitrary $M$, i.e., the optimal number of active users is approximately $Mu_{ASE}^*$.\footnote{In fact, $Mu_{ASE}^*$ need to be rounded as the number of active users can only be integers.}  Additionally, the lower-bound of ASE $\underline{T}$ with optimal $\underline{K}_{ASE}^*$ is
\begin{equation}\label{eqtopt}
\underline{T}^{opt}=\lambda_bMG(u_{ASE}^*).
\end{equation}

We can see that the lower-bound of ASE will increase linearly with the number of antennas. We  define $G(u_{ASE}^*)$ as the gain on ASE per antenna (GAPA), which means the gain on ASE by deploying one more antenna. The next procedure is to maximize  $G(u)$ to find the value of $u_{ASE}^*$ and GAPA. We have the following results.
\newtheorem{lemma}{Lemma}
\begin{lemma}\label{usolution}$u_{ASE}^*$ is the solution of the equation $\int_0^\infty  {\frac{{1}}{z}} \frac{{1 - e^{ - z\frac{1}{u} + z}  - z\frac{1}{u}e^{ - z\frac{1}{u} + z} }}{{e^{ - z}  + z^{\frac{2}{\alpha }} \gamma \left( {1 - \frac{2}{\alpha },z} \right)}}dz=0$. There is a unique solution for this equation, which satisfies $u\in[0,1]$ and can be derived through the bisection method.
\end{lemma}

\begin{IEEEproof}
See Appendix \ref{proofu}.
\end{IEEEproof}

Based on Lemma \ref{usolution}, we can derive $u_{ASE}^*$ and $G(u_{ASE}^*)$.

\subsubsection{Numerical Illustrations}

\begin{figure}[!t]
\begin{minipage}[t]{0.5\linewidth}
\centering
\includegraphics[width=3.2in]{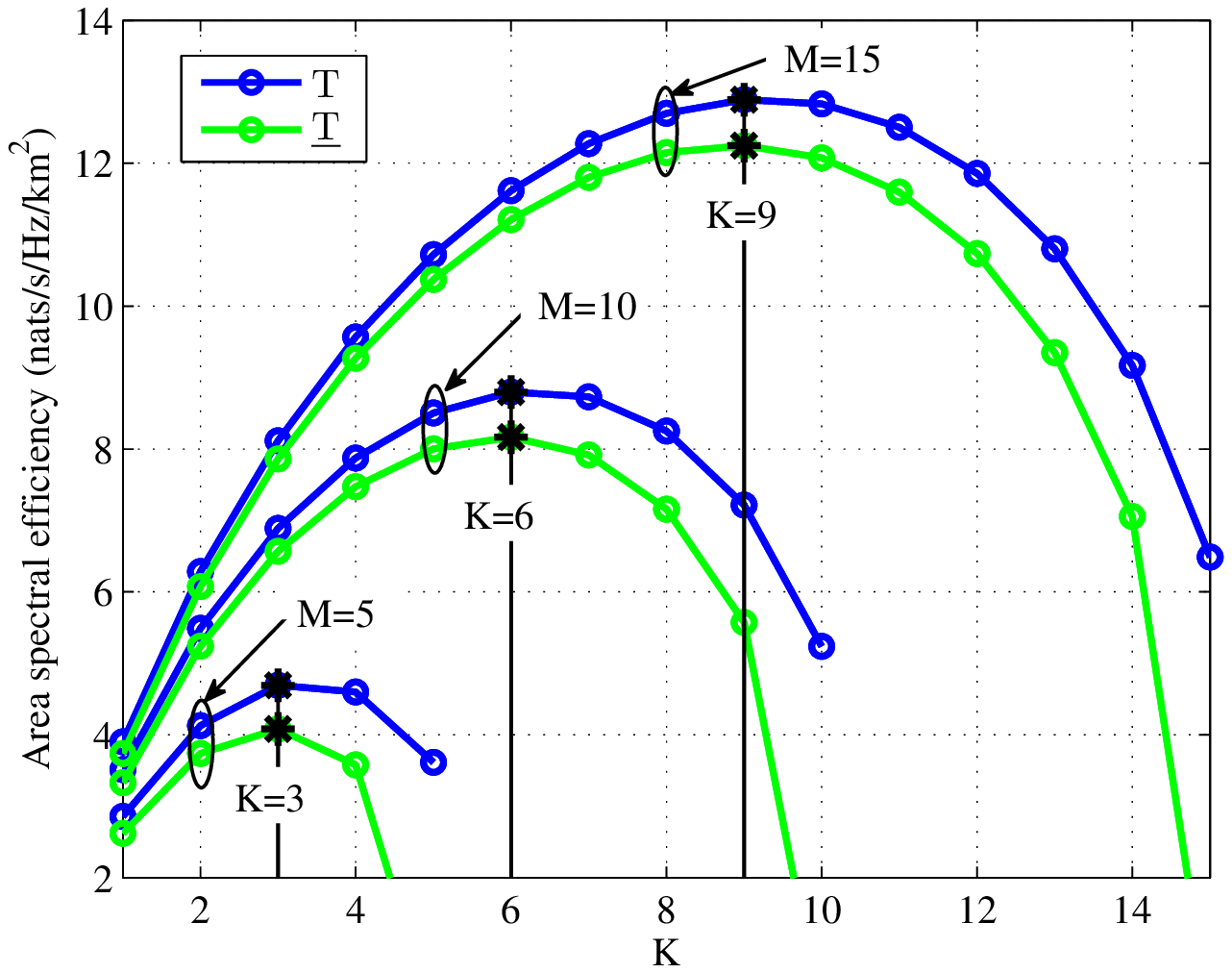}
\caption{ASE with different $K$. $\lambda_b=1km^{-2}$. $\alpha=4$.}
\label{optimalk}
\end{minipage}
\hspace{1ex}
\begin{minipage}[t]{0.5\linewidth}
\centering
\includegraphics[width=3.2in]{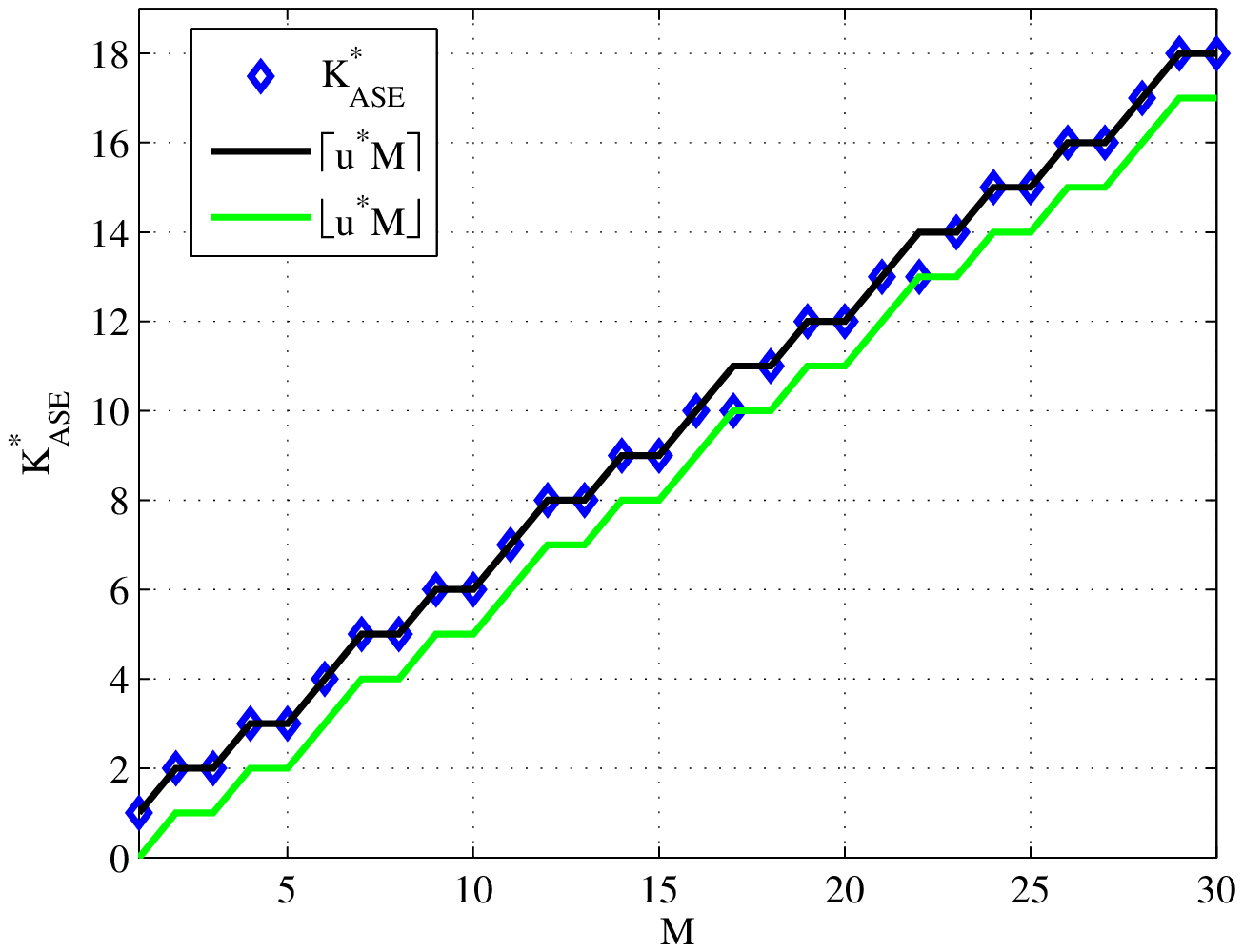}
\caption{Optimal number of active users. $\alpha=4$.}
\label{optimalum}
\end{minipage}
\end{figure}

In simulations, we set the path loss exponent $\alpha=4$. Following Lemma \ref{usolution}, we derive $u_{ASE}^*=0.5913$ and the GAPA $G(u_{ASE}^*)=0.8165 \ nats/s/Hz$. Thus, given the number of antennas $M$, the optimal number of active users to maximize the lower-bound of ASE is approximately $0.5913M$.\footnote{This value needs to be rounded  as the number of active users can only be integers.} Fig. \ref{optimalk} depicts $T$ and $\underline{T}$ with different $K$. Firstly, we can find that both $T$ and $\underline{T}$ are concave with respect to $K$, which demonstrates Proposition \ref{prok}. For $M=5,10,15$, the optimal number of active users that maximizes $T$ and $\underline{T}$  are both $K_{ASE}^*=\underline{K}_{ASE}^*=3,6,9$. This demonstrates our analysis about the optimal $\underline{K}_{ASE}^*$. Furthermore, the conclusion based on $\underline{T}$ is also true for the  expression $T$. To further demonstrate this, we obtain the optimal $K_{ASE}^*$ that maximizes $T$ for different $M$ by exhaustive search. The optimal $K_{ASE}^*$ with different $M$ is provided in Fig. \ref{optimalum}. Although $\underline{K}_{ASE}^*=u_{ASE}^*M$ is derived based on the lower-bound $\underline{T}$, when considering the expression $T$, we find that the optimal $K_{ASE}^*$ is either $\lfloor u_{ASE}^*M\rfloor$ or $\lceil u_{ASE}^*M \rceil$. This demonstrates that our analysis based on the lower-bound can be applied to the original expression directly.

%


\begin{figure}[!t]
\centering
\includegraphics[width=3.2in]{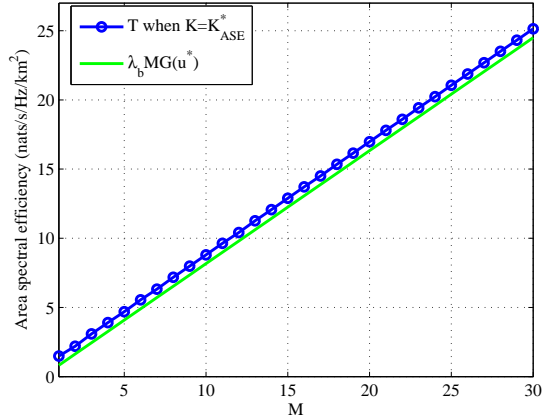}
\caption{ASE with $M$ considering optimal $K^*$. $\lambda_b=1km^{-2}$. $\alpha=4$.}
\label{throughputm}
\end{figure}

Interestingly, in subsection A, we found that the lower-bound of ASE is a concave function of $M$ when $K$ is fixed. However, if the number of active users $K$ is set as $u_{ASE}^*M$, from $(\ref{eqtopt})$, we know that the lower-bound of ASE approximately increases linearly with the number of antennas. When $K$ set as the optimal number, Fig. \ref{throughputm} illustrates $T$ and $\underline{T}$ with different $M$. We find that both  $T$ and $\underline{T}$  increase linearly with $M$. \footnote{Part of similar phenomenons were observed via numerical results in \cite{li:sdma2}. However, the reasons behind these phenomenons have not been discovered. Instead of simulations, we arrive at these results through rigorous theoretical analysis. Thus, the reasons behind have been fully explained.} That is to say, with optimal configuration, increasing the number of antennas is as efficient as increasing the BS density, i.e., both of them can boost the network capacity linearly.  Actually, from Theorem \ref{thlowbound}, it is not difficult to find that, if only the  ratio $u=\frac{K}{M}$ remains fixed, ASE can increase linearly with the number of antennas. The difference is that, if we set $u=u_{ASE}^*$, ASE will be maximized. The reasons are as follows. Recalling Fig. \ref{datarateu} in Section \ref{secanalysis}, the average data rate of each active user mainly depends on $u$. That is to say, although we increase $M$, if only $u$ is fixed, the average data rate of each active user nearly remains fixed, i.e., the increase in the desired power is counter-balanced by the channel power of the interfering links. Furthermore, the number of active users $K=uM$, which increases linearly with the number of antennas. Therefore, we arrive at the conclusion that ASE increases linearly with the number of antennas.


%
%
%

\section{Energy Optimal System Parameters}\label{secoptimal}
In this section, we pursue an energy optimal strategy to minimize network energy consumption while guaranteeing the ASE design target. Specifically, we formulate an optimization problem where the BS density, the number of antennas and active users are jointly optimized. The optimal algorithm and a suboptimal algorithm are proposed to solve the non-convex problem.

\subsection{Problem Formulation}
While guaranteeing the network ASE is above the target $T^{tar}$, we aim at minimizing the average network energy consumption per unit area, i.e., $\lambda_b\left(\frac{P}{\eta}+MP_c+K^3P_{pre}+P_0\right)$. We try to answer the fundamental question how many BSs and antennas should be deployed and how many users should be scheduled in each slot. Therefore,  three parameters $\left(\lambda_b,M,K\right)$ will be jointly optimized. We can formulate the energy minimization problem as

\setlength{\arraycolsep}{0.0em}
\begin{eqnarray}\label{problenergy}
 \mathbf{P2:}  \quad &    \min \limits_{\lambda_b>0;M,K\in \bf{N}^+} \quad  & \lambda_b\left(\frac{P}{\eta}+MP_c+K^3P_{pre}+P_0\right)    \qquad \nonumber \\
&s.t. &  \lambda_b K E\left[R\right] \ge T^{tar},     \\
 && K \le M.                           \nonumber
\end{eqnarray}
\setlength{\arraycolsep}{5pt}

It is intuitive that, the inequality constraint $\lambda_b K E\left[R\right] \ge T^{tar}$ can be replaced by the equality constraint $\lambda_b K E\left[R\right] = T^{tar}$. Thus, we have the relationship
\begin{equation}\label{eqdensitym}
\lambda_b=\frac{T^{tar}}{K E\left[R\right]}.
\end{equation}
Plugging $(\ref{eqdensitym})$ into $\mathbf{P2}$, we can arrive at  the following equivalent optimization problem,

\setlength{\arraycolsep}{0.0em}
\begin{eqnarray}\label{problenergy1}
   \max \limits_{M,K\in \bf{N}^+} &  \qquad & \frac{KE\left[R\right]}{\frac{P}{\eta}+MP_c+K^3P_{pre}+P_0} \nonumber \\
s.t. & \qquad &  K \le M.
\end{eqnarray}
\setlength{\arraycolsep}{5pt}

In fact, the problem is energy-efficiency maximization problem. In most cases, energy consumption minimization problem is not equivalent to the energy-efficiency maximization problem. However, due to the equality $\lambda_b=\frac{T^{tar}}{KE[R]}$, the problem $(\ref{problenergy})$$(\ref{problenergy1})$ in this paper are equivalent. In other words, the optimal combination of the number of antennas and active users that maximizes the energy-efficiency is  also the optimal solution that minimizes the network energy consumption.  Although the optimization problem $\mathbf{P2}$ is non-convex, the optimal algorithm and a suboptimal algorithm will be proposed  to solve $\mathbf{P2}$ in this section.

\subsection{Optimal Algorithm}

We first relax $M,K$ to $(0,+\infty)$ in optimization problem $(\ref{problenergy1})$. Thus, we arrive at a fractional programming problem. The major obstacle to solving this problem is the complicated expression of $E\left[R\right]$, especially the complex relationship between $E\left[R\right]$ and $K$. This motivates us to study the optimal $M$ when $K$ is fixed, which is more tractable. From Proposition \ref{m}, we know that $E\left[R\right]$ is a concave function of $M$. As the denominator is a linear function, the above optimization problem is a concave fractional program, and the objective function  is pseudoconcave.  Due to the generalized concavity of the objective function, we have the following results about the concave fractional programs \cite{fp:schaible},
\begin{enumerate}
  \item A local maximum is the global maximum;
  \item It is possible to solve concave fractional programs with standard concave programming algorithms.
\end{enumerate}

Based on the first order derivative of the objective function, we obtain the optimal $M_{EC}^*$.

\begin{theorem}\label{thm}
When $K$ is given, the optimal number of antenna $M_{EC}^*=\mathbf{round}\left(\max \left(\widetilde{M},K\right)\right)$,\footnote{The operation $\mathbf{round}(\cdot)$ chooses the integer which leads to a smaller network energy consumption from $\{\lfloor \cdot \rfloor,\lceil \cdot \rceil\}$.} where $\widetilde{M}$ is the solution of the equation
\begin{equation}\label{eqsolution}
\underbrace{\frac{\partial E\left[R\right]}{\partial M}\left({\frac{P}{\eta}+MP_c+K^3P_{pre}+P_0}\right)-E\left[R\right]P_c}_{F(M)}=0.
\end{equation}
The above equation has a unique solution, which is located in $[K-1,+\infty)$ and can be derived through the bisection method.
\end{theorem}

\begin{IEEEproof}
See Appendix \ref{proofom}
\end{IEEEproof}

From Theorem \ref{thm}, we can find how $M_{EC}^*$ varies with other system parameters. Based on the method of implicit differentiation, we have $\frac{\partial M_{EC}^*}{\partial\left(\frac{P/\eta + K^3P_{pre} + P_0}{P_c}\right)}= - \frac{P_c\frac{\partial E\left[R\right]}{\partial M}}{\frac{\partial ^2 E\left[R\right]}{\partial M^2}\left({\frac{P}{\eta}+MP_c+K^3P_{pre}+P_0}\right)}>0.$ Thus, $M_{EC}^*$ increases with the ratio $\frac{P/\eta + K^3P_{pre} + P_0}{P_c}$. In other words, the smaller the circuit power per antenna $P_c$ compared with other energy consumption is, the more antennas are preferred.

Based on Theorem \ref{thm}, we have derived the optimal number of antennas $M_{EC}^*$ when given $K$. Therefore, by comparing all possible results for $K$, we can derive the optimal algorithm for $\mathbf{P2}$.
\begin{itemize}
  \item First, obtain the optimal number of antennas $M_{EC}^*$ for all possible $K\in \{1,2,3,...\}$ based on Theorem \ref{thm}.
  \item Then, substitute $M_{EC}^*$ into $(\ref{eqdensitym})$, we can derive $\lambda_{bEC}^*$.
  \item Select the optimal solution $\left(\lambda_{bEC}^*, M_{EC}^*,K_{EC}^*\right)$ that minimizes the network energy consumption.
\end{itemize}

\subsection{Suboptimal Algorithm Based on The Lower-Bound}
Although the optimal algorithm proposed in this subsection can derive the optimal solution of $\mathbf{P2}$, the  exhaustive search for $K_{EC}^*$ is time consuming.  To reduce the computational complexity, we also propose a suboptimal algorithm based the lower-bound in Theorem \ref{thlowbound}. Replacing $E\left[R\right]$ with $\underline{E\left[R\right]}$ and relaxing $M,K$ to $(0,+\infty)$ in  $(\ref{problenergy1})$, we arrive at the following related but not equivalent problem.

\setlength{\arraycolsep}{0.0em}
\begin{eqnarray}\label{problenergy2}
   \max \limits_{M,K\in (0,\infty)}  & \qquad & \frac{K\underline{E\left[R\right]}}{\frac{P}{\eta}+MP_c+K^3P_{pre}+P_0} \nonumber \\
s.t.& \qquad &   K \le M.
\end{eqnarray}
\setlength{\arraycolsep}{5pt}

In Section \ref{secimpact}, we have demonstrated that the optimal number of active users that maximizes $K\underline{E[R]}$ is $u_{ASE}^*M$. Based on this result, for the above problem, the optimal $\underline{M}_{EC}^*$ and $\underline{K}_{EC}^*$ satisfy the following properties.

\begin{lemma}\label{lemmaconstraint}
The optimal $(\underline{M}_{EC}^*,\underline{K}_{EC}^*)$ for problem $(\ref{problenergy2})$ must satisfy $\underline{K}_{EC}^* \le u_{ASE}^*\underline{M}_{EC}^*$, i.e., the constraint $K \le M$ can be removed.
\end{lemma}

\begin{IEEEproof}
From the analysis in Section \ref{secanalysis}, we know that $K\underline{E\left[R\right]}$ is maximized at $u_{ASE}^*\underline{M}_{EC}^*$ when given $\underline{M}_{EC}^*$. Hence, for arbitrary $K'>u_{ASE}^*\underline{M}_{EC}^*$, we have $\left.K\underline{E\left[R\right]}\right|_{K=K'}<\left.K\underline{E\left[R\right]}\right|_{K=u_{ASE}^*\underline{M}_{EC}^*}$. Besides, the denominator $\left.\frac{P}{\eta}+MP_c+K^3P_{pre}+P_0\right|_{K=K'}>\left.\frac{P}{\eta}+MP_c+K^3P_{pre}+P_0\right|_{K=u_{ASE}^*\underline{M}_{EC}^*}$. Above all, we have $\left.\frac{K\underline{E\left[R\right]}}{\frac{P}{\eta}+MP_c+K^3P_{pre}+P_0}\right|_{K=K'} < \left.\frac{K\underline{E\left[R\right]}}{\frac{P}{\eta}+MP_c+K^3P_{pre}+P_0}\right|_{K=u_{ASE}^*\underline{M}_{EC}^*}$. Therefore, the optimal $(\underline{M}_{EC}^*,\underline{K}_{EC}^*)$  must satisfy $\underline{K}_{EC}^* \le u_{ASE}^*\underline{M}_{EC}^*$.
\end{IEEEproof}

From Theorem \ref{thmk}, we know that the numerator $K\underline{E\left[R\right]}$ is concave with respect to $(M,K)$. Besides, it is not difficult to show that the denominator is convex with respect to $(M,K)$. Hence, the above optimization problem is a concave fractional program, and the objective function  is pseudoconcave.  We use alternating optimization to solve the optimization problem $(\ref{problenergy2})$, i.e., $M,K$ are optimized sequentially. About the optimal $\underline{M}_{EC}^*,\underline{K}_{EC}^*$, we have the following results.

\begin{theorem}\label{thoptk}
When $M$ is fixed, the optimal $\underline{K}_{EC}^*$ for   $(\ref{problenergy2})$ satisfies $\underline{F}_K(M,K)=0$, where
\begin{equation}
\begin{array}{l}
\underline{F}_K(M,K)   =\frac{\partial \left( {K\underline {E\left[ R \right]} } \right) }{{\partial K}}\left( \frac{P}{\eta } + MP_c+ K^3 P_{pre}  + P_0  \right) -3K^3 \underline {E\left[ R \right]} P_{pre}.
\end{array}
\end{equation}
The unique solution for the above equation lies in $(0,u_{ASE}^*M)$ and can be obtained through the bisection method.
\end{theorem}

\begin{IEEEproof}
See Appendix \ref{proofsk}.
\end{IEEEproof}

Also based on the method of implicit differentiation, we have
\begin{equation}
\frac{\partial \underline{K}_{EC}^*}{\partial\left(\frac{P/\eta + MP_c +P_0}{P_{pre}}\right)}=-\frac{P_{pre}\frac{\partial \left( {K\underline {E\left[ R \right]} } \right) }{{\partial K}}}{\frac{{\partial ^2 \left( {K\underline {E\left[ R \right]} } \right)}}{{\partial K^2 }}\left( {\frac{P}{\eta } + MP_c  + K^3 P_{pre}  + P_0 } \right) - 6K^2E\left[R\right]P_{pre}}.
\end{equation}
From Lemma \ref{lemmaconstraint}, we know $\underline{K}_{EC}^*\le u^*M$. Therefore, we have $\frac{\partial \left( {K\underline {E\left[ R \right]} } \right) }{{\partial K}}>0$ for $\underline{K}_{EC}^*$. The denominator of above equation is demonstrated be to negative in Appendix \ref{proofsk}. Above all, $\frac{\partial \underline{K}_{EC}^*}{\partial\left(\frac{P/\eta + MP_c +P_0}{P_{pre}}\right)}>0$, i.e., the optimal $\underline{K}_{EC}^*$ increases with the ratio $\frac{P/\eta + MP_c +P_0}{P_{pre}}$. That is to say,  compared with other energy consumption, the smaller the energy consumption for precoding is, the more active users are preferred.

\begin{theorem}\label{thoptm}
When $K$ is fixed, the optimal $\underline{M}_{EC}^*$ for $(\ref{problenergy2})$ satisfies the following equation
\begin{equation}
\underbrace{{\frac{{\partial \underline {E\left[ R \right]} }}{{\partial M}}\left( {\frac{P}{\eta } + MP_c  + K^3 P_{pre}  + P_0 } \right) - \underline {E\left[ R \right]} P_c}}_{\underline{F}_M(M,K)}=0.
\end{equation}
There is a unique solution for this equation, which lies in $(K,+\infty)$. Additionally, the solution can be derived through the bisection method.
\end{theorem}

\begin{IEEEproof}
See Appendix \ref{proofsm}.
\end{IEEEproof}

Similar to the discussion for Theorem \ref{thm}, we can arrive at the conclusion that $\underline{M}_{EC}^*$ increases with the ratio $\frac{P/\eta + K^3P_{pre} + P_0}{P_c}$, which means the property of  $\underline{M}_{EC}^*$  is consistent with $M_{EC}^*$.

In summary, the optimal $\underline{M}_{EC}^*,\underline{K}_{EC}^*$ can be derived via Theorem \ref{thoptk}, \ref{thoptm} when the other parameter is fixed. Due to the generalized concavity of the objective function, the optimal solution for  $(\ref{problenergy2})$ can be derived  by alternating optimization. Above all, we get the suboptimal algorithm for $\mathbf{P2}$ as follows.\footnote{As the optimization in each phase is non-decreasing, alternating optimization can converge to local optimal point. For the quasiconcave function, we can derive the global optimal point.}
\begin{itemize}
  \item Optimize $K$ by using Theorem \ref{thoptk};
  \item Update $M$ by using Theorem \ref{thoptm};
  \item Repeat the above two procedures until convergence. Round the results, we obtain the suboptimal $\underline{M}_{EC}^*,\underline{K}_{EC}^*$;
  \item Substitute $\underline{M}_{EC}^*,\underline{K}_{EC}^*$ into $(\ref{eqdensitym})$, we can obtain the suboptimal $\underline{\lambda}_{bEC}^*$.
\end{itemize}

For complexity analysis, we denote the search region for $K_{EC}^*, M_{EC}^*, \underline{K}_{EC}^*,\underline{M}_{EC}^*$  as $[1,L]$. For the optimization of $M_{EC}^*, \underline{K}_{EC}^*,\underline{M}_{EC}^*$ based on Theorem \ref{thm},\ref{thoptk},\ref{thoptm}, the complexity of bisection method is $O\left(\log _2L\right)$. As the optimal algorithm needs exhaustive search for $K_{EC}^*$, the complexity of the optimal algorithm is $O\left(L\log _2 L\right)$. Additionally, the complexity of the suboptimal algorithm is $O\left(N\log _2 L\right)$, where $N$ is the number of iterations. Through the suboptimal algorithm, the complexity can be reduced by $\frac{L}{N}$ times. Indeed, $M$ is unbounded in the optimization problems. We need to set $L$ sufficiently large (e.g., 100) in simulations. On the other hand, our numerical results will show that the suboptimal algorithm can converge to the stable point with small number of iterations  ($N<5$). Therefore, the complexity can be reduced sufficiently.

\subsection{Numerical Illustrations}

In this subsection, we demonstrate the proposed algorithms through numerical results. We set the path loss exponent $\alpha=4$. The parameters of BS energy consumption model are consistent with \cite{zhang:model}. We consider three types of BSs, i.e., macro BSs, micro BSs, and pico BSs. For the macro BSs, $P=54dBm$, $\eta=0.388$, $P_c=16.9W$, $P_{pre}=1.74W$ $P_0=65.8$. For the micro BSs, $P=46dBm$, $\eta=0.285$, $P_c=13.3W$, $P_{pre}=1.74W$ $P_0=65.8W$. For the pico BSs, $P=33dBm$, $\eta=0.08$, $P_c=6.8W$, $P_{pre}=1.74W$ $P_0=1.5W$. To study the potential gains of MU-MIMO, we consider two compared scenarios: (i) SU-MIMO, this corresponds to the scenario only single-user MIMO is applied, i.e., $K=1$; (ii) Single Antenna, which means only single antenna BSs are considered.\footnote{For the scenario SU-MIMO, $K$ is set as 1 and $M$ is optimized based on Theorem \ref{thm}. For the scenario Single Antenna, we set $M=K=1$.}

\begin{figure}[!t]
\begin{minipage}[t]{0.5\linewidth}
\centering
\includegraphics[width=3.2in]{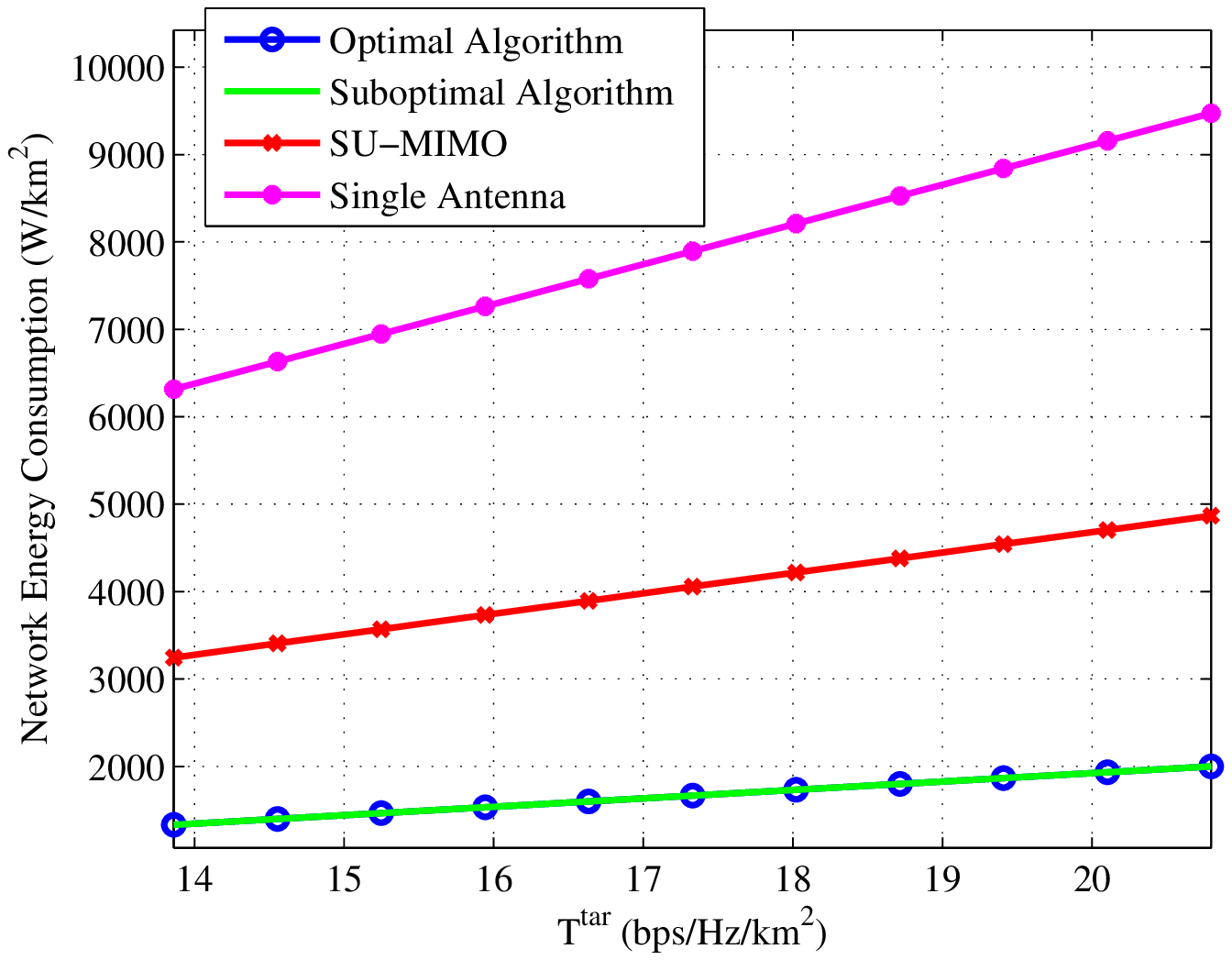}
\caption{Network energy consumption for macro BSs.}
\label{54dBmEC}
\end{minipage}
\hspace{1ex}
\begin{minipage}[t]{0.5\linewidth}
\centering
\includegraphics[width=3.2in]{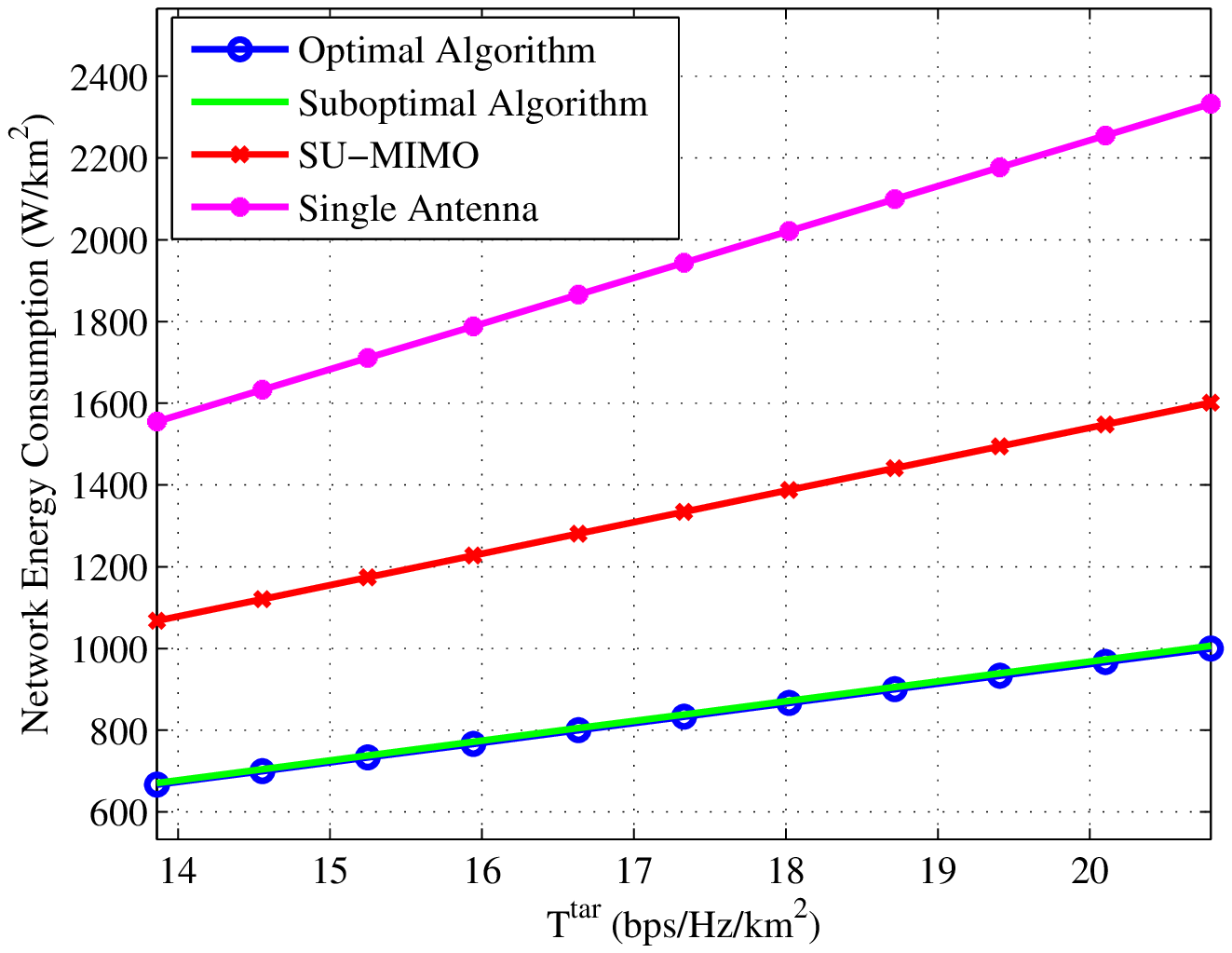}
\caption{Network energy consumption for micro BSs.}
\label{46dBmEC}
\end{minipage}
\end{figure}

%

\begin{figure}[!t]
\begin{minipage}[t]{0.5\linewidth}
\centering
\includegraphics[width=3.2in]{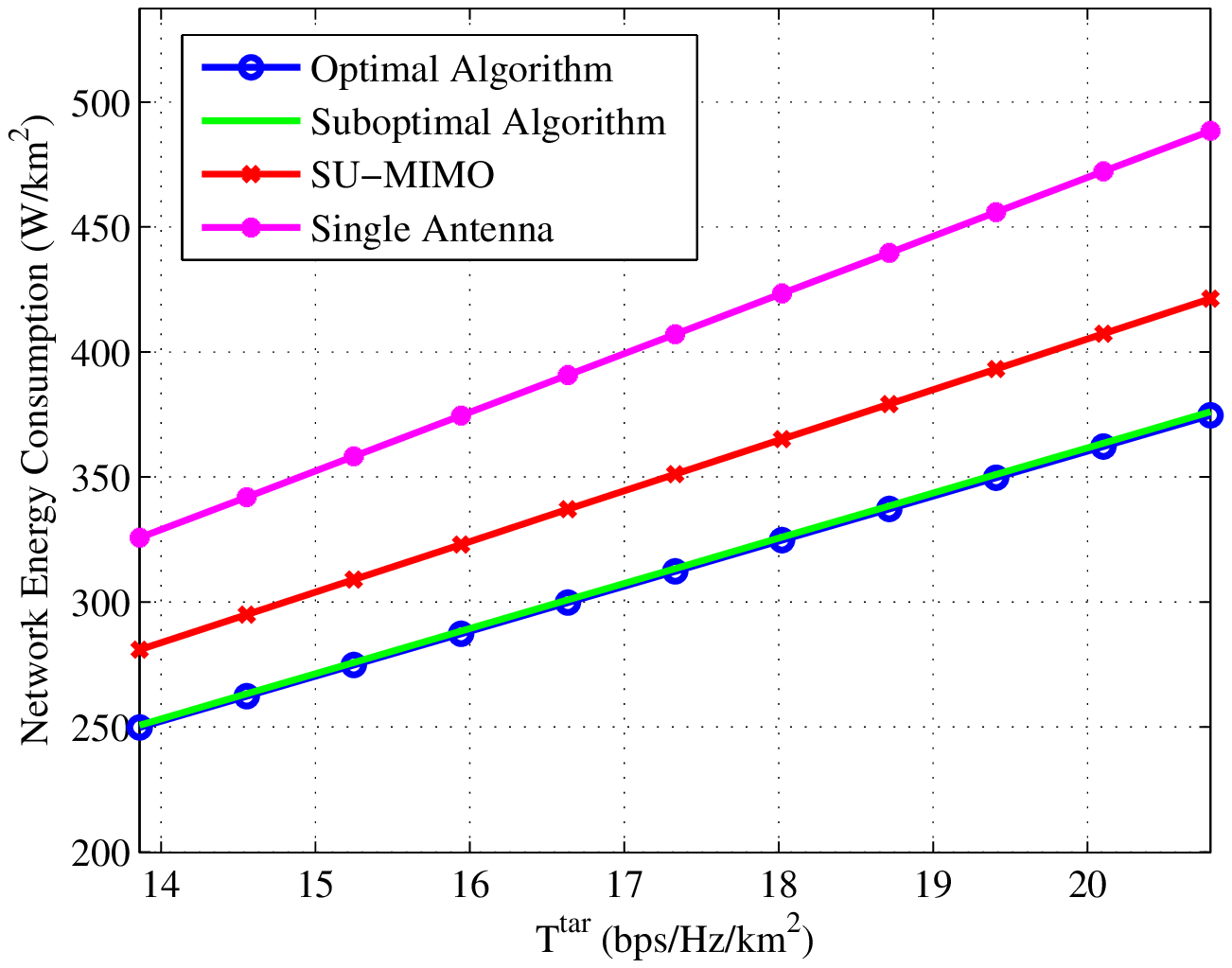}
\caption{Network energy consumption for pico BSs.}
\label{33dBmEC}
\end{minipage}
\hspace{1ex}
\begin{minipage}[t]{0.5\linewidth}
\centering
\includegraphics[width=3.2in]{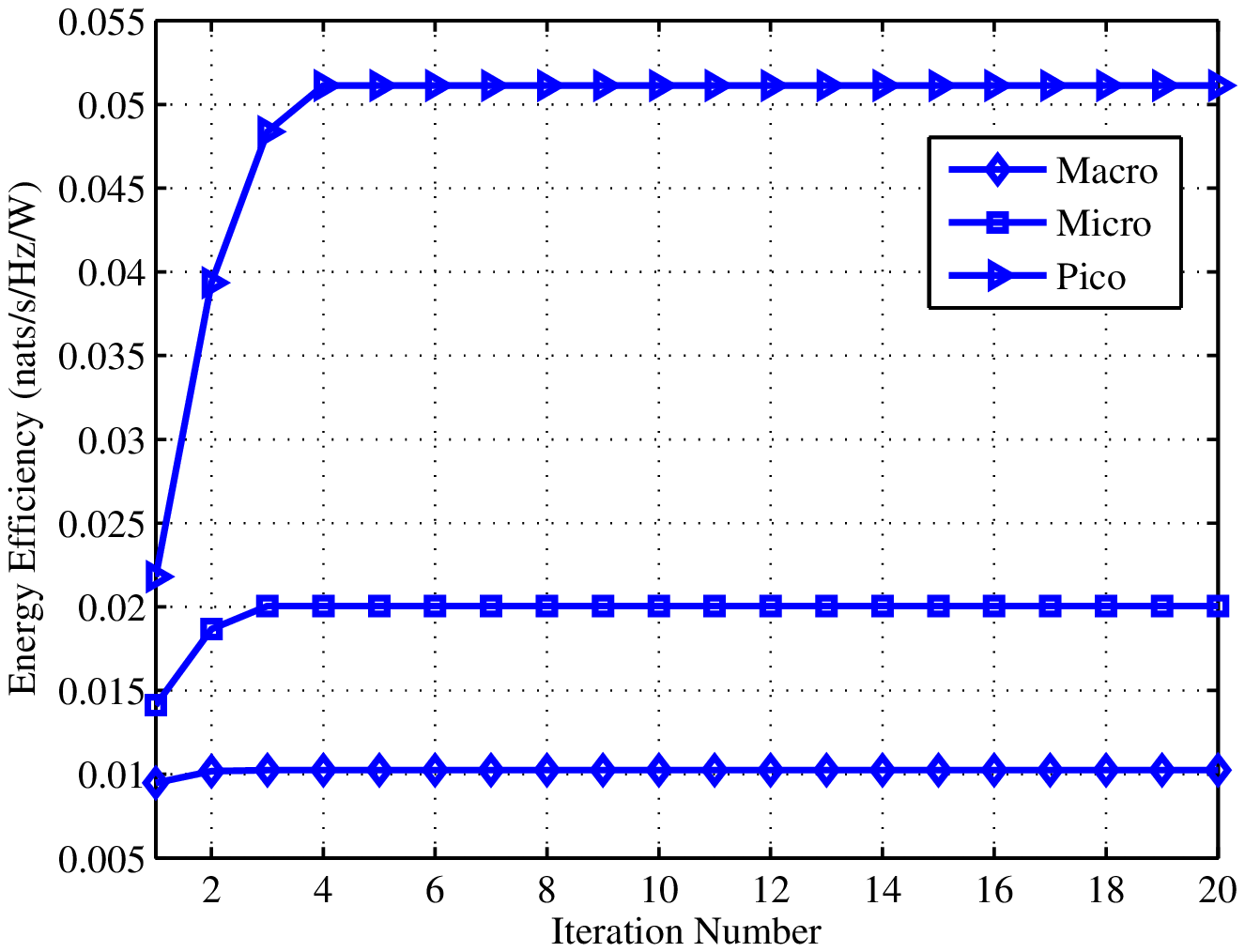}
\caption{Convergence of the suboptimal algorithm.}
\label{convergence}
\end{minipage}
\end{figure}

%

Fig. \ref{54dBmEC}-\ref{33dBmEC} depict the network energy consumption considering the configurations of macro, micro, and pico BSs. We can easily find that the suboptimal algorithm can achieve near-optimal performance. Compared to the single-antenna scenario, if we only consider single-user transmission, we find that deploying multi-antenna can save 50\%, 33\%, 14\% energy for macro, micro, and pico networks, respectively. If MU-MIMO is considered, 28\%, 27\%, 10\% more energy can be further saved for the scenarios of  macro, micro, and pico networks, respectively. That is to say, equipping multi-antenna with larger BSs can save more energy. Additionally, we find that the network energy consumption increases linearly with the ASE target $T^{tar}$. The reason is that, the optimal $M_{EC}^*,K_{EC}^*$ is the solution that maximizes the energy-efficiency, which is unrelated to $T^{tar}$. Thus, the optimal network energy consumption can be expressed as
\begin{equation}
\begin{array}{l}
NEC=\frac{T^{tar}}{ \left.KE\left[R\right]\right|_{M=M_{EC}^*,K=K_{EC}^*}}\left(\frac{P}{\eta}+M_{EC}^*P_c+K_{EC}^{*3}P_{pre}+P_0\right). \quad
\end{array}
\end{equation}
Therefore, we know that the network energy consumption increases linearly with the $T^{tar}$.  Through simulation results, we find that $M_{EC}^*=35,K_{EC}^*=6$ for macro BSs; $M_{EC}^*=11,K_{EC}^*=3$ for micro BSs, and $M_{EC}^*=5,K_{EC}^*=2$ for pico BSs. That is to say, it is preferable to equip more antennas with larger BSs which have higher transmit power. And more users will be scheduled for larger BSs. The reasons are as follows, recalling the discussion for the optimal $M_{EC}^*, \underline{K}_{EC}^* ,\underline{M}_{EC}^*$ in Theorem \ref{thm},\ref{thoptk},\ref{thoptm},   the smaller $P_c$ and $P_{pre}$ compared with the total energy consumption, the larger $M_{EC}^*, \underline{K}_{EC}^*, \underline{M}_{EC}^*$ is. For macro, micro, pico BSs, the ratio between $\frac{P}{\eta}+P_0$ and $P_c$ is 39.03, 11.42, 3.88, respectively. And ratio between  $\frac{P}{\eta}+P_0$ and $P_{pre}$ is 379.13, 87.3, 15.19. This means $P_c$ and $P_{pre}$ occupy entirely different fractions of total energy consumption for different type of BSs, i.e., the smallest for the macro BSs, the largest for pico BSs. Thus, we observe the fact that more antennas and active users are preferred for larger BSs.

Fig. \ref{convergence} depicts the convergence of  the suboptimal algorithm. We find that the suboptimal algorithm can converge to the stable point with small iteration number. For configurations of  macro, micro, and pico BSs,  the iteration number is less than 5.  This demonstrates the efficiency of the suboptimal algorithm. Moreover, we find the maximum energy-efficiency for macro, micro and pico BSs are approximately 0.01, 0.02, and 0.05 $nats/s/Hz/W$, respectively. This means deploying small BSs is an efficient way to improve the network energy-efficiency.


\section{Conclusions}\label{secconclusion}
By modeling the positions of BSs as PPP, this paper has studied two fundamental questions. The first one is the functional relationship between ASE and other system parameters.  Based on the expression and a tight lower-bound of the average data rate of the typical user, we have found that average data rate of the typical user is mainly related to the ratio between the number of active users and the number of antennas. Additionally, ASE is jointly concave with respect to the number of antennas and the number of active users. For the purpose of maximizing ASE, we have demonstrated that the number of active users should be set as a fixed portion of the number of antennas. Furthermore, with the number of active users set to optimal, we have found that ASE increases linearly with the number of antennas.

Another fundamental question  we try to answer is the optimal combination of the BS density, the number of antennas and active users to minimize the network energy consumption. We have discovered that the optimal number of antennas and active users is the solution that maximizes energy-efficiency. Numerical results have demonstrated that the proposed suboptimal algorithm can achieve near optimal performance. Besides, we also found that  it is preferable to equip more antennas and schedule more active users for the BSs which have higher transmit power.

Using stochastic geometry, we have fully explored the performance of ASE and network energy consumption of the cellular network with multi-antenna BSs.  The insightful results in this paper shed many lights on system design. Especially, we have found that ASE increases linearly with the number of antennas when considering optimal configuration. Besides, we have obtained GAPA to  accurately quantify of the gains by deploying more antennas. Furthermore, the energy optimal deployment strategy has also been proposed. How to densify the network in an energy-efficient way is studied. Further extension of this work is to consider the cooperation between BSs, where more gains will be predicted since the inter-cell interference is handled.

\appendices

\section{}\label{proofase}
\begin{IEEEproof}[Proof of Theorem \ref{thase}]
Denote the distance between the typical  user and the serving BS $\|x_0\|$ as $r_0$, and the distance between the typical user and the interfering BSs $\|x_i\|$ as $r_i,i>0$. From the null probability of PPP, the distance between the typical user  and the serving BS is expressed as  $f(r_0)=2\pi \lambda_b r_0 e^{ - \pi \lambda_b r_0^2 }$. Following the law of total expectation, the average data of the typical user can be calculated as
\begin{equation}\label{eqasecal}
\begin{array}{l}
E\left[R\right]=E_{r_0}\left[E_{\Phi_b,\mathbf{g}}\left[\left.\log \left(1+\frac{g_{00}r_0^{-\alpha}}{\sum \limits_{x_i \in \Phi_b \backslash \{x_0\} }g_{i0}r_i^{-\alpha}}\right)\right| r_0\right]\right].
\end{array}
\end{equation}
On the basis of Lemma 1 in \cite{hamdi:lemma}, the conditional expectation can be derived as follows.
\begin{equation}
\begin{array}{l}
 E_{\Phi_b,\mathbf{g}}\left[\left.\log \left(1+\frac{g_{00}r_0^{-\alpha}}{\sum_{x_i \in \Phi_b \backslash \{x_0\} }g_{i0}r_i^{-\alpha}}\right)\right| r_0\right]           \\
  = \int_0^\infty  {\frac{ {1 - E_{g_{00}}\left[ {e^{ - zg_{00} } } \right]}}{z}E_{\Phi_b,\mathbf{g}}\left[ {e^{ - z\sum_{x_i  \in \Phi \backslash \{ x_0\} } {g_{i0} r_i^{ - \alpha } r_0^\alpha  } } } \right]dz}     \\
  \stackrel{\left(a\right)}{=} \int_0^\infty  {\frac{1 - \left( {\frac{1}{{1 + z}}} \right)^{M + 1 - K}}{z}E_{\Phi_b}\left[ \prod_{x_i  \in \Phi_b \backslash \{ x_0 \} } {\frac{1}{{\left( {1 + zr_i^{ - \alpha } r_0^\alpha  } \right)^K }}} \right]dz}     \\
  \stackrel{\left(b\right)}{=} \int_0^\infty  {\frac{1 - \left( {\frac{1}{{1 + z}}} \right)^{M + 1 - K}}{z}e^{- 2\lambda_b \pi \int_{r_0 }^\infty  {\left( {1 - \frac{1}{{\left( {1 + zy^{ - \alpha } r_0^\alpha  } \right)^K }}} \right)ydy} }dz}   \\
 = \int_0^\infty  {\frac{1 - \left( {\frac{1}{{1 + z}}} \right)^{M + 1 - K}}{z}e^{- \lambda_b \pi z^{\frac{2}{\alpha }} r_0^2 \int_{z^{ - \frac{2}{\alpha }} }^\infty  {\left( {1 - \left( \frac{1}{1 + u^{ - \frac{\alpha }{2}} } \right)^{ K} } \right)du}}dz},
 \end{array}
\end{equation}
where $(a)$ follow from the fact that $g_{00},g_{i0}$ are Gamma distributed, $(b)$ follows from the probability generating functional (PGFL) of PPP \cite{book:baccelli}. Substituting the conditional expectation, we have the following results.
\begin{equation}
\begin{array}{l}
E_{r_0}\left[E_{\Phi_b,\mathbf{g}}\left[\left.\log \left(1+\frac{g_{00}r_0^{-\alpha}}{\sum_{x_i \in \Phi_b \backslash \{x_0\} }g_{i0}r_i^{-\alpha}}\right)\right| r_0\right]\right]   \\
= \int_0^\infty  \int_0^\infty  \frac{{\left( {1 - \left( {\frac{1}{{1 + z}}} \right)^{M + 1 - K} } \right)}}{z}2\pi \lambda_b r_0 e^{ - \pi \lambda_b r_0^2 }       e^{ - \lambda_b \pi z^{\frac{2}{\alpha }} r_0^2 \int_{z^{ - \frac{2}{\alpha }} }^\infty  {\left( {1 - \frac{1}{{\left( {1 + u^{ - \alpha /2} } \right)^K }}} \right)du} }  dz dr_0  \\
= \int_0^\infty  \frac{ {\frac{{1}}{z}} \left({1 - \left( {\frac{1}{{1 + z}}} \right)^{M + 1 - K} }\right)}{{\left( {1 + z} \right)^{ - K}  + z^{\frac{2}{\alpha }} KB\left( {\frac{z}{{1 + z}},1 - \frac{2}{\alpha},K + \frac{2}{\alpha} } \right)}}dz.
 \end{array}
\end{equation}
Two major algebraic manipulations in the last step are: i) reversing the order of integration, ii) integration by parts to simplify the expression.
\end{IEEEproof}

\section{}\label{prooflowebound}
\begin{IEEEproof}[Proof of Theorem \ref{thlowbound}]
The lower-bound is the result by applying the Jensen's inequality  $E\left[ \log \left(1+\frac{1}{x}\right)\right] \ge \log \left(1+\frac{1}{E\left[x\right]}\right)$. Applying this inequality, we have the following results.
\begin{equation}
\begin{array}{l}
 E_{\Phi _b ,{\bf{g}}} \left[ {\log \left( {1 + \frac{{g_{00} r_0^{ - \alpha } }}{{\sum_{x_i  \in \Phi _b \backslash \{ x_0 \} } {g_{i0} r_i^{ - \alpha } } }}} \right)} \right] \\
  = E_{\Phi _b ,{\bf{g}}} \left[ {\log \left( {1 + \frac{1}{{\sum_{x_i  \in \Phi _b \backslash \{ x_0 \} } {\frac{{g_{i0} }}{{g_{00} }}r_i^{ - \alpha } r_0^\alpha  } }}} \right)} \right] \\
  \ge E_{\Phi _b } \left[ {\log \left( {1 + \frac{1}{{E_{\bf{g}} \left[ {\sum_{x_i  \in \Phi _b \backslash \{ x_0 \} } {\frac{{g_{i0} }}{{g_{00} }}r_i^{ - \alpha } r_0^\alpha  } } \right]}}} \right)} \right] \\
  \stackrel{\left(a\right)}{=}\underbrace{E_{\Phi _b } \left[ {\log \left( {1 + \frac{{\frac{{M - K}}{K}}}{{\sum_{x_i  \in \Phi _b \backslash \{ x_0 \} } {r_i^{ - \alpha } r_0^\alpha  } }}} \right)} \right]}_{\underline{E\left[R\right]}} , \\
 \end{array}
\end{equation}
where $(a)$ following from $g_{i0},i\ge0$ are independent and Gamma distributed. Specifically, $E_{\bf{g}}\left[ {\sum_{x_i  \in \Phi _b \backslash \{ x_0 \} } {\frac{{g_{i0} }}{{g_{00} }}r_i^{ - \alpha } r_0^\alpha  } } \right]={E\left[\frac{1}{g_{00} }\right]}\sum_{x_i  \in \Phi _b \backslash \{ x_0 \} } {{E\left[{g_{i0} }\right]}r_i^{ - \alpha } r_0^\alpha  }$. As $g_{i0} \sim Gamma (K,1)$ for $i>0$, we have $E\left[g_{i0}\right]=K$. According to \cite{cook:inversegamma}, $\frac{1}{g_{00}}$ follows the inverse gamma ditribution $IG(M+K-1,1)$. Furthermore, the expectation $E\left[\frac{1}{g_{00}}\right]=\frac{1}{M-K}$. Hence, we arrive at above results.

The major procedures of the calculation for $\underline{E\left[R\right]}$ is similar to the proof of Theorem \ref{thase}. Firstly, we need to obtain the conditional expectation
\begin{equation}
\begin{array}{l}
 E_{\Phi _b } \left[ {\left. {\log \left( {1 + \frac{{\frac{{M - K}}{K}}}{{\sum {_{x_i  \in \Phi _b \backslash \{ x\} } } r_i^{ - \alpha } r_0^\alpha  }}} \right)} \right|r_0 } \right] \\
  = \int_0^\infty  {\frac{1 - e^{ - z\frac{{M - K}}{K}} }{z}} E\left[ {e^{ - z\sum {_{x_i  \in \Phi _b \backslash \{ x\} } } r_i^{ - \alpha } r_0^\alpha  } } \right]dz \\
  = \int_0^\infty  {\frac{1 - e^{ - z\frac{{M - K}}{K}} }{z}} e^{ - \lambda _b \pi z^{\frac{2}{\alpha }} r_0^2 \int_{z^{ - \frac{2}{\alpha }} }^\infty  {\left( {1 - e^{ - y^{ - \frac{\alpha }{2}} } } \right)dy} } dz.  \\
 \end{array}
\end{equation}
Then, following the law of total expectation, we have
\begin{equation}
\begin{array}{l}
 E_{\Phi_b}  \left[ {\log \left( {1 + \frac{{\frac{{M - K}}{K}}}{{\sum\limits_{x_i  \in \Phi \backslash \{ x\} } {x_i^{ - \alpha } x_0^\alpha  } }}} \right)} \right] \\
  = \int_0^\infty  \int_0^\infty  \frac{{1 - e^{ - z\frac{{M - K}}{K}} }}{z}e^{ - \lambda _b \pi z^{\frac{2}{\alpha }} x_0^2 \int_{z^{ - \frac{2}{\alpha }} }^\infty  {\left( {1 - e^{ - y^{ - \frac{\alpha }{2}} } } \right)dy} }  2\pi \lambda _b r_0 \exp \left( { - \pi \lambda _b r_0^2 } \right)dz dr_0  \\
  = \int_0^\infty  {\frac{{1}}{z}} \frac{{1 - e^{ - z\frac{{M - K}}{K}} }}{{ e^{ - z}  + z^{\frac{2}{\alpha }} \gamma \left( {1 - \frac{2}{\alpha },z} \right)}}dz. \\
 \end{array}
\end{equation}

\end{IEEEproof}

\section{}\label{proofconcave}
\begin{IEEEproof}[Proof of Theorem \ref{thmk}]
Based on  $(\ref{eqtmk})$, considering integration is a linear operation, we only need to prove that ${K\left({1 - e^{ - z\frac{{M - K}}{K}} }\right)}$ is concave. The corresponding  Hessian matrix is expressed as $H=\frac{z^2}{K}e^{-z\frac{M}{K}+z}\left[\begin{array}{cc} -1&\frac{M}{K}\\ \frac{M}{K}&-\frac{M^2}{K^2}\end{array}\right].$ For arbitrary non-zero vector $\mathbf{y}=[y_1,y_2]^T\in \mathbf{R}^2$,  we have
\begin{equation}
\begin{array}{ll}
\mathbf{y}^TH\mathbf{y} & = \frac{z^2}{K}e^{-z\frac{M}{K}+z}[y_1,y_2]\left[
\begin{array}{cc}
-1&\frac{M}{K}\\
\frac{M}{K}&-\frac{M^2}{K^2}
\end{array}
\right]
\left[
\begin{array}{c}
y_1\\
y_2
\end{array}
\right]                 \\
&=-\frac{z^2}{K}e^{-z\frac{M}{K}+z}\left(y_1-\frac{M}{K}y_2\right)^2    \\
&<0.
\end{array}
\end{equation}
Hence, $H$ is negative-definite. Therefore,  we arrive at Theorem \ref{thmk}.
\end{IEEEproof}

\section{}\label{proofu}
\begin{IEEEproof}[Proof of Lemma \ref{usolution}]
The first and second order derivatives of  $G(u)$  are as follows,
\begin{equation}
\begin{array}{l}
 \frac{\partial }{{\partial t}}G\left( u \right) = \int_0^\infty  {\frac{{1}}{z}} \frac{{1 - e^{ - z\frac{1}{u} + z}  - z\frac{1}{u}e^{ - z\frac{1}{u} + z} }}{{e^{ - z}  + z^{\frac{2}{\alpha }} \gamma \left( {1 - \frac{2}{\alpha },z} \right)}}dz,
 \frac{{\partial ^2 }}{{\partial t^2 }}G\left( u \right) =  - \int_0^\infty  {\frac{{1}}{z}} \frac{{z^2 \frac{1}{{u^3 }}e^{ - z\frac{1}{u} + z} }}{{ e^{ - z}  + z^{\frac{2}{\alpha }} \gamma \left( {1 - \frac{2}{\alpha },z} \right)}}dz. \\
 \end{array}
\end{equation}
As $\frac{{\partial ^2 }}{{\partial t^2 }}G\left( u \right)<0$,  $G(u)$ is a concave function of $u$. Thus, if ignoring the constraint $u\in [0,1]$, the optimal $u_{ASE}^*$ could be derive by setting the first order derivative as zero, i.e., $\frac{\partial }{{\partial t}}G\left( u \right)=0$.
As $\frac{{\partial ^2 }}{{\partial t^2 }}G\left( u \right)<0$, $\frac{\partial }{{\partial t}}G\left( u \right)$ is a decreasing function of $u$. For $\frac{\partial }{{\partial t}}G\left( 0 \right)$ and $\frac{\partial }{{\partial t}}G\left( 1 \right)$, we have the following results. First, for $u=0$, we have $ \lim _{u \to 0} \left( {1 - e^{ - z\frac{1}{u} + z}  - z\frac{1}{u}e^{ - z\frac{1}{u} + z} } \right) = \lim _{u \to 0} \left( {1 - z\frac{{e^z }}{{\frac{{e^{z\left( {1/u} \right)} }}{{1/u}}}}} \right) = \lim _{u \to 0} \left( {1 - \frac{{e^z }}{{e^{z\left( {1/u} \right)} }}} \right)= 1.$ Thus, $\frac{\partial }{{\partial t}}G\left( 0 \right)= \int_0^\infty  {\frac{{1}}{z}} \frac{1}{{ e^{ - z}  + z^{\frac{2}{\alpha }} \gamma \left( {1 - \frac{2}{\alpha },z} \right)}}dz >0$. For $u=1$, $\frac{\partial }{{\partial t}}G\left( 1 \right) =  - \int_0^\infty  {\frac{{1}}{{ e^{ - z}  + z^{\frac{2}{\alpha }} \gamma \left( {1 - \frac{2}{\alpha },z} \right)}}dz}<0.$ Therefore, we know that there is a unique solution for $\frac{\partial }{{\partial t}}G\left( u \right)=0$ in $[0,1]$. The solution can be derived via the bisection method.
\end{IEEEproof}

\section{}\label{proofom}
\begin{IEEEproof}[Proof of Theorem \ref{thm}]When ignoring the constraint $K \le M$, the local optimal $M$ can be obtained by setting the first order derivative of the objective function in $(\ref{problenergy1})$ as zero. Therefore, we arrive at $(\ref{eqsolution})$.
The first order derivative of the left item of the equation $(\ref{eqsolution})$ is
\begin{equation}
\frac{\partial F(M)}{\partial M} = \frac{\partial ^2 E\left[R\right]}{\partial M^2}\left({\frac{P}{\eta}+MP_c+K^3P_{pre}+P_0}\right).
\end{equation}
From Proposition \ref{m}, we know that $\frac{\partial ^2 E\left[R\right]}{\partial M^2}<0$. Thus, we have $\frac{\partial F(M)}{\partial M}<0$, i.e., $F(M)$ is a decreasing function of $M$.
For $K-1$, $F(K-1)=\left.\frac{\partial E\left[R\right]}{\partial M}\right|_{M=K-1}\left({\frac{P}{\eta}+(K-1)P_c+K^3P_{pre}+P_0}\right)>0$. If $M \to +\infty$, the first order derivative  $\lim _{M \to  + \infty } \frac{{\partial E[R]}}{{\partial M}}=0$. Thus, $\lim _{M \to \infty}F(M)=-\lim_{M\to +\infty}E\left[R\right]P_c<0$. Combining the fact that $F(M)$ is a decreasing function of $M$, we know that there is a unique  solution for $(\ref{eqsolution})$, which is located in $[K-1, +\infty)$. And the solution can derived through the bisection method. Because the solution is unique, the solution is the global optimal point. Above all, considering the constraint $K \le M$ and the fact that $M$ is an integer, we get Theorem \ref{thm}.
\end{IEEEproof}

\section{}\label{proofsk}
\begin{IEEEproof}[Proof of Theorem \ref{thoptk}]
When $M$ is fixed, we derive the optimal $\underline{K}_{EC}^*$ by  setting the first order derivative of the objective function in $(\ref{problenergy2})$ as zero. By this,  we have the  equation $\underline{F}_K(M,K)=0$. The first order derivative $\frac{\partial }{{\partial K}}{\underline{F}_K(M,K)} = \frac{{\partial ^2 \left( {K\underline {E\left[ R \right]} } \right)}}{{\partial K^2 }}\left( {\frac{P}{\eta } + MP_c  + K^3 P_{pre}  + P_0 } \right) - 6K^2E\left[R\right]P_{pre}$.  As $K\underline {E\left[ R \right]} $ is concave with respect to $K$, we have $\frac{\partial }{{\partial K}}{\underline{F}_K(M,K)}<0$. Therefore, $\underline{F}_K(M,K)$ is decreasing function of $K$.
For $K=0$, $\underline{F}_K(M,K)=
{\left.\frac{\partial }{{\partial K}}\left( {K\underline {E\left[ R \right]} } \right)\right|_{K=0}\left( {\frac{P}{\eta } + MP_c  + P_0 } \right)}>0$. Combining Lemma \ref{lemmaconstraint}, we know that the solution for $\underline{F}_K(M,K)=0$ lies in $(0,u_{ASE}^*M)$ and can be obtained through the bisection method.
\end{IEEEproof}

\section{}\label{proofsm}
\begin{IEEEproof}[Proof of Theorem \ref{thoptm}]
Similar to Theorem \ref{thoptk}, the optimal $\underline{K}_{EC}^*$  is obtained by  setting the first order derivative of the objective function in $(\ref{problenergy2})$ as zero. Thus, we arrive at the equation $\underline{F}_M(M,K)=0$. The first order derivative  $\frac{\partial }{{\partial M}}\underline{F}_M(M,K) = \frac{{\partial \underline {^2 E\left[ R \right]} }}{{\partial M^2 }}\left( {\frac{P}{\eta } + MP_c  + K^3 P_{pre}  + P_0 } \right)<0$.  For $M=K$, $\underline{F}_M(M,K)=\left.\frac{{\partial \underline {E\left[ R \right]} }}{{\partial M}}\right|_{M=K}\left( {\frac{P}{\eta } + KP_c  + K^3 P_{pre}  + P_0 } \right)>0$. For $M\to +\infty$, we know that $\lim_{M\to +\infty}\underline{F}_M(M,K)=-\lim_{M\to +\infty}\underline{E\left[R\right]}P_c<0$. Thus, there is a unique solution for $\underline{F}_M(M,K)=0$, which lies in $(K,+\infty)$. Additionally, the solution can be derived through the bisection method.
\end{IEEEproof}

\ifCLASSOPTIONcaptionsoff
  \newpage
\fi

\end{document}